\documentclass[twocolumn]{aastex7}
\usepackage{multirow}
\usepackage{booktabs}
\usepackage{graphicx}
\usepackage{amsmath}
\usepackage{ulem}

\begin{document}

\title{A Correlation Between the Final Separation and Mass Ratio from Common Envelope Simulations}

\author[orcid=0000-0003-4050-9920,sname='Borges']{Sarah V. Borges}
\affiliation{Department of Physics and Astronomy, University of Wisconsin-Milwaukee, 3135 North Maryland Avenue, Milwaukee, WI 53211, USA}
\email[show]{svborges@uwm.com}  

\begin{abstract}

Analytical models for common envelope evolution (CEE), particularly the energy formalism, are used in binary population synthesis to predict post-CEE configurations. This formalism is based on an efficiency parameter $\alpha$, which relates the orbital energy released during CEE to that required to unbind the envelope of the giant. However, one of the main challenges is that CEE is a multiscale, multiphysics process. As a result, there may not be a universal value for $\alpha$, or even a general expression. Using 13 3D simulations of CEE with RGBs (1 and 2 M$_\odot$ primary; four mass ratios; with and without corotation), we present an empirical linear correlation between the post-plunge-in separation and the mass ratio, normalized by the giant radius. This trend for the plunge-in phase of CEE persists across RGB, AGB, and supergiant simulations in the literature, even for partially bound envelopes. Therefore, $\alpha$ from simulations should not be used to predict the final separation, but rather as a diagnostic of whether sufficient orbital energy has been liberated to completely eject the envelope immediately after the radial plunge. If this condition is not met, further in-spiral is expected in later stages of CEE, which may explain why the final separation of post-CEE observations is generally smaller than those predicted by the linear fit. Our results reinforce the idea that a better description could emerge if CEE is treated as a sequence of distinct phases, rather than treating it as a single event governed by $\alpha$.

\end{abstract}


\section{Introduction}

Common envelope evolution (CEE) is a short phase of binary evolution where the two stars are within the same stellar envelope. The resulting drag forces lead to the loss of orbital energy and angular momentum, producing close binary systems. CEE is therefore essential for the formation rate of systems and events such as X-ray binaries, Type Ia supernovae, and gravitational wave sources from merging compact objects~\citep[for a review, see][]{2020cee..book.....I}.

Despite its significance, CEE remains one of the least understood phases of binary evolution, primarily due to the lack of direct observations. While the luminous red nova~(LNRe) V1309 Sco~\citep{2011A&A...528A.114T} provides a glimpse of CEE in action, it results in a merger and is not fully representative of classical CEE, where both stars survive the interaction. As a result, hydrodynamical simulations are a key tool for studying this phase. However, these simulations are computationally expensive and can only model a limited number of systems at a time. Thus, for population synthesis studies, the analytical energy formalism is commonly used. This formalism assumes that the loss of orbital energy is responsible for ejecting the giant envelope, with an efficiency parameter $\alpha$ that ranges from 0 (no energy transfer) to 1 (perfect transfer). In some cases, $\alpha$ may exceed unity if additional sources, such as recombination energy, contribute to the unbinding of the envelope.

The challenge of predicting $\alpha$ highlights a fundamental issue with this formalism. Observational studies of post-CEE systems yield conflicting results: some suggest that $\alpha$ remains roughly constant across systems~\citep{2010A&A...520A..86Z}, while others indicate that it depends on parameters such as the mass ratio~\citep{2011MNRAS.411.2277D}. Moreover, different types of systems appear to favor different values of $\alpha$. For instance, close detached white dwarf~(WD) + brown dwarf binaries tend to require low values \citep[$\alpha < 0.4$,][]{2022MNRAS.513.3587Z}, whereas high-mass X-ray binaries suggest significantly higher values \citep[$\alpha > 0.8$,][]{2014MNRAS.442.1980Z}. 

The significance of determining $\alpha$ lies in its ability to predict the final configuration of a system based on its initial parameters~\citep{2019MNRAS.490.2550I}. However, recent simulations suggest that we may be able to skip $\alpha$ in favor of an empirical relationship. For example, \cite{2016MNRAS.460.3992N} simulated pre-double WDs CEE systems and found a relationship between the final orbital period, the initial orbital period, and the stellar masses. More recently, \cite{2020A&A...644A..60S} performed CEE simulations with a low-mass AGB giant and found that the final-to-initial orbital separation appears to vary linearly with the mass ratio.

Motivated by the findings of \cite{2020A&A...644A..60S}, we carried out new simulations of low-mass red giant branch~(RGB) stars and found that these systems follow a similar linear relationship between final separation and mass ratio, despite differences in giant structure. This paper is organized as follows: in Sec.~\ref{sec:methods}, we describe our simulation setup and present the results. In Sec.~\ref{sec:other_simulations}, we compare our results with other CEE simulations.  In Sec.~\ref{sec:formalism}, we compare the correlation with the predictions of the energy formalism. In Sec.~\ref{sec:observations}, we confront the simulations with observational data. In Sec.~\ref{sec:moving}, we discuss our results. Finally, in Sec.~\ref{sec:summary}, we summarize our findings.

\section{CEE simulations}\label{sec:methods}

\subsection{Methods}

For the CEE simulations, we used \texttt{MANGA}~\citep{2017MNRAS.471.3577C}, a moving-mesh module for the SPH code \texttt{ChaNGa}~\citep{jetley2008,jetley2010,2015ComAC...2....1M}. Currently, \texttt{MANGA}~includes radiation hydrodynamics~\citep{2020MNRAS.493.5397C}, general relativistic hydrodynamics~\citep{2020MNRAS.496..206C}, moving-boundary conditions~\citep{2020MNRAS.494.4616P}, and magnetohydrodynamics (Prust \& Chang, in preparation). \texttt{MANGA}~has previously been used to simulate CEE~\citep{2019MNRAS.486.5809P,2020MNRAS.494.4616P,2023MNRAS.526.5365V}, as well as stellar mergers~\citep{2017MNRAS.471.3577C}, and tidal disruption events~\citep{2021MNRAS.501.1748S}.

For the initial conditions, we evolved the stars using MESA \citep{2011ApJS..192....3P,2013ApJS..208....4P,2015ApJS..220...15P,2018ApJS..234...34P,2019ApJS..243...10P,2023ApJS..265...15J} from the pre-main sequence to the RGB phase. We studied two different primary stars. First, a $1~\text{M}_\odot$ RGB with a radius of 78~R$_{\odot}$. Second, a $2~\text{M}_\odot$ RGB with a radius of 53~R$_{\odot}$. Next, we constructed a star with an entropy profile that matched that of the MESA star. To achieve this, we used an adiabatic equation of state ($\gamma = 5/3$), which means that recombination energy was not included in the unbinding of the envelope. The core of the star was replaced with a point particle with a softening potential, and the envelope's radial profiles of density and temperature were mapped onto an unstructured particle mesh. The companion is also modeled as a point particle and is positioned at the surface of the RGB.  We consider the $1~\mathrm{M}_\odot$ giant to be in corotation with the binary orbit, while for the $2~\mathrm{M}_\odot$ case we simulate with and without corotation. \added{The initial orbit is circular.}

\begin{table}
\centering
\caption{ Parameters of our CEE simulations. Each set shares the same primary mass $M_{1,\text{t}}$, core mass $M_{1,\text{c}}$, and primary radius~$R_{1}$. The mass ratio is $q = M_2/M_{1,\text{t}}$ and $a_f$ is the semi-major axis at the end of the plunge. Unless specified, the simulations consider corotation and uses the \texttt{HLLC} Riemann solver.}
\begin{tabular}{ccccc}
\hline
$M_{1,\text{t}}$~($\text{M}_\odot$) & $M_{1,\text{c}}$~($\text{M}_\odot$) & $R_{1}$~($\text{R}_\odot$) & $q$ & $a_{\rm f}$~($\text{R}_\odot$) \\
\hline
\multirow{4}{*}{1} & \multirow{4}{*}{0.38} & \multirow{4}{*}{78} & 0.05 & 3.5 \\
                   &                       &                     & 0.3  & 7.8 \\
                   &                       &                     & 0.6  & 14.4 \\
                   &                       &                     & 0.9  & 22.7 \\
\hline
\multirow{4}{*}{2} & \multirow{4}{*}{0.37} & \multirow{4}{*}{53} & 0.25 & 3.5 \\
                   &                       &                     & 0.5  & 6.9 \\
                   &                       &                     & 0.75 & 11.2 \\
                   &                       &                     & 1.0  & 16.7 \\

\hline
\multirow{4}{*}{2$^{\dagger}$} & \multirow{4}{*}{0.37} & \multirow{4}{*}{53} & 0.25 & 2.8 \\
                   &                       &                     & 0.5  & 5.7 \\
                   &                       &                     & 0.75 & 9.1 \\
                   &                       &                     & 1.0  & 12.4 \\

\hline
\multirow{1}{*}{2$^{+}$} & \multirow{1}{*}{0.37} & \multirow{1}{*}{53} &  0.5  & 3.8 \\
\hline

\end{tabular}
\label{tab:our_simulations}
Notes: $^{\dagger}$non-corotation; $^{+}$ \texttt{HLL} Riemann solver.
\end{table}

Our simulations follow the same procedure described by \citet{2019MNRAS.486.5809P} and \citet{2023MNRAS.526.5365V}, incorporating only a few modifications. First, while their work used the \texttt{HLL} Riemann solver, we opted for the \texttt{HLLC} solver. In all cases, the background density was set to $10^{-20}$g\,cm$^{-3}$. The simulations evolved for $800$ days. To test the impact of the Riemann solver, we also ran one additional model with the \texttt{HLL} solver (the $2~\mathrm{M}_\odot$ RGB, $q=0.5$ case). Table~\ref{tab:our_simulations} summarizes the parameters of our 13 simulations. Finally, the total number of particles differs between models: for the 1~M$_\odot$ RGB, we used 1.6 million particles in total, of which 1 million represent the star; for the 2~M$_\odot$ RGB, we used 2 million particles, with 1.2 million for the star. \added{The softening radius of the companion is given by $(M_2/\mathrm{M}_\odot)\mathrm{R}_\odot$ and for the cores, it is 1~$\mathrm{R}_\odot$ for the 1~$\mathrm{M}_\odot$ RGB, and 0.5~$\mathrm{R}_\odot$ for the 2~$\mathrm{M}_\odot$ RGB. }

\subsection{Results}

\begin{figure*}
\centering
\includegraphics[width=0.47\textwidth]{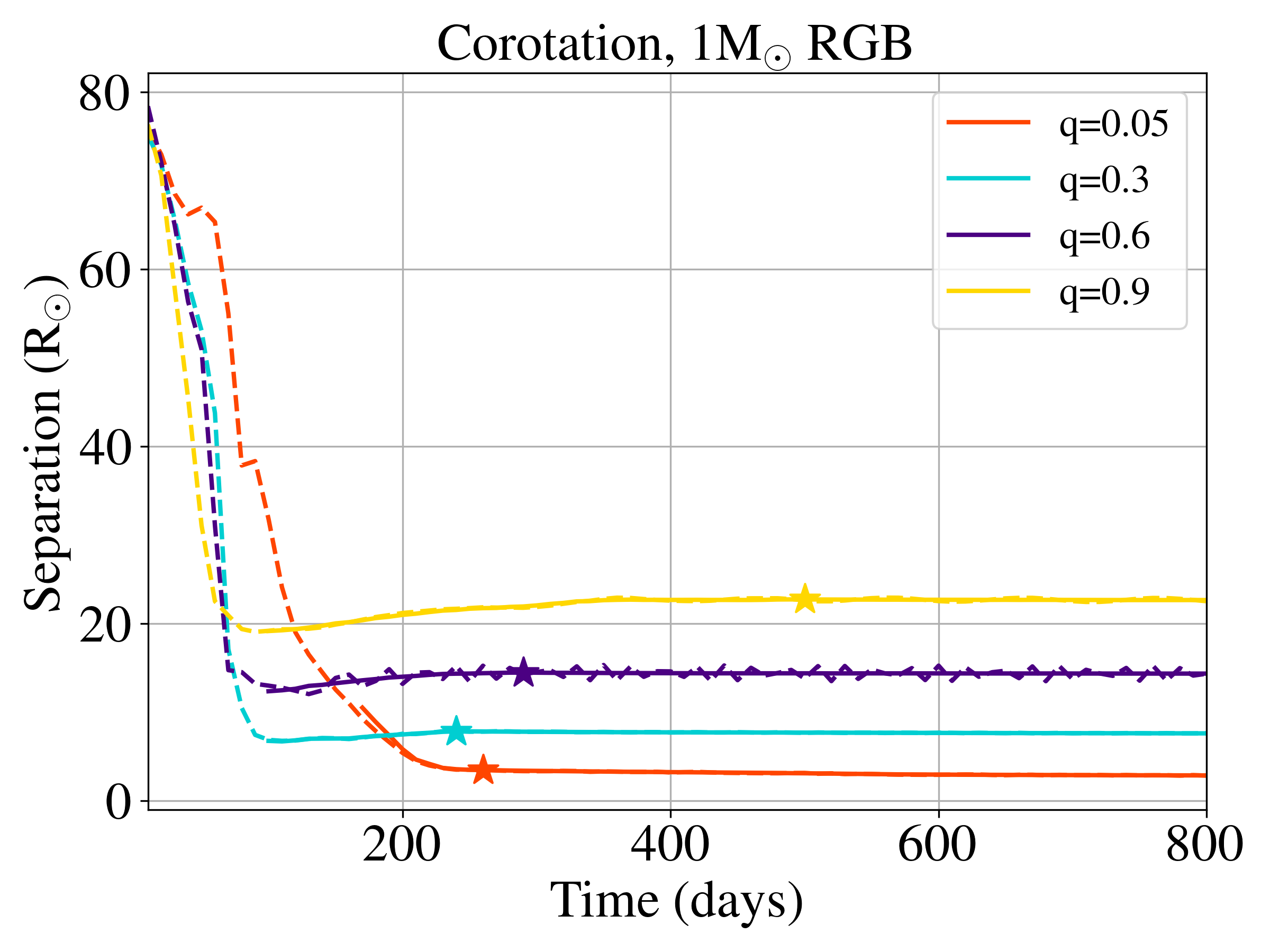}
\includegraphics[width=0.47\textwidth]{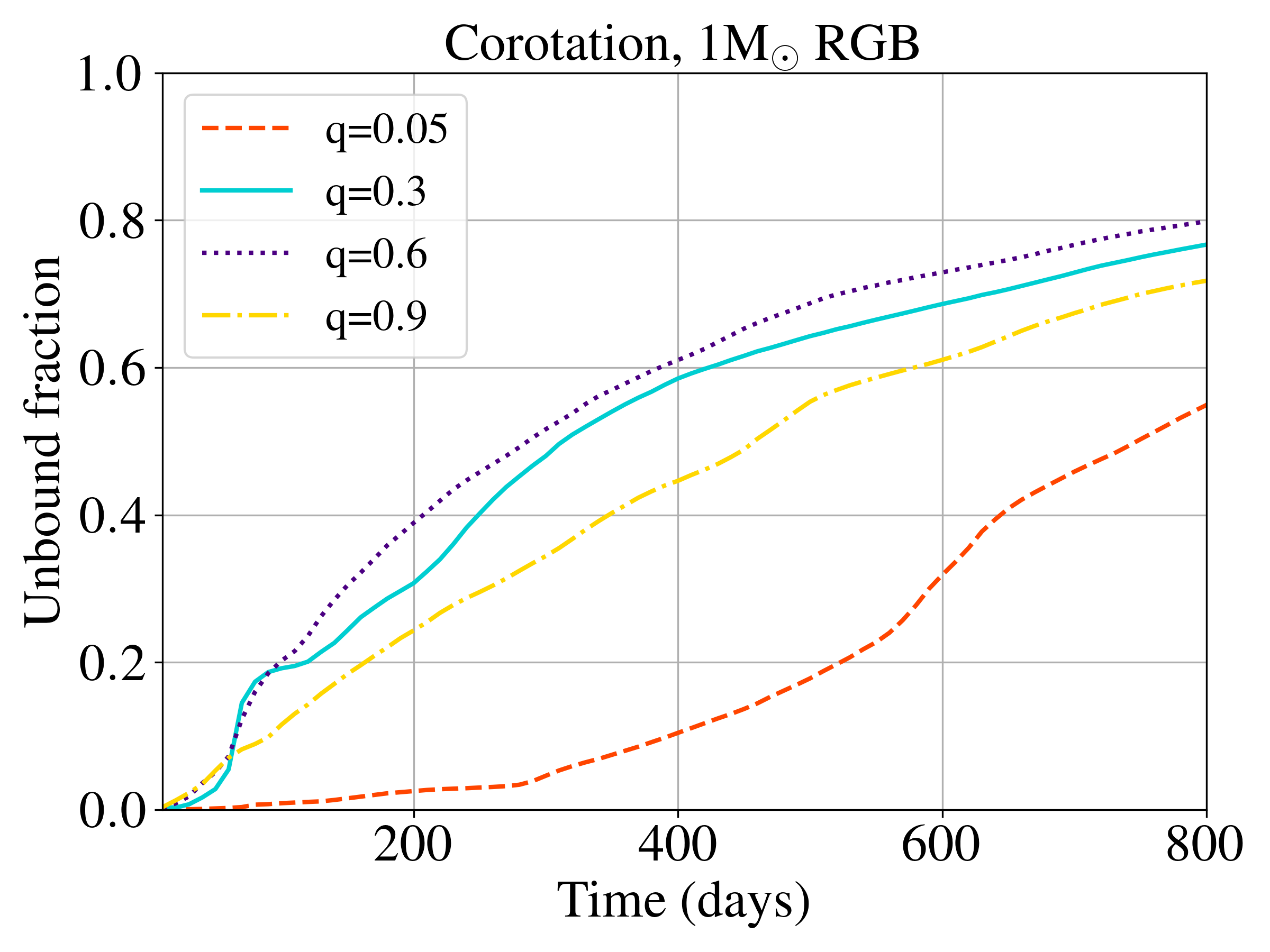}
\includegraphics[width=0.47\textwidth]{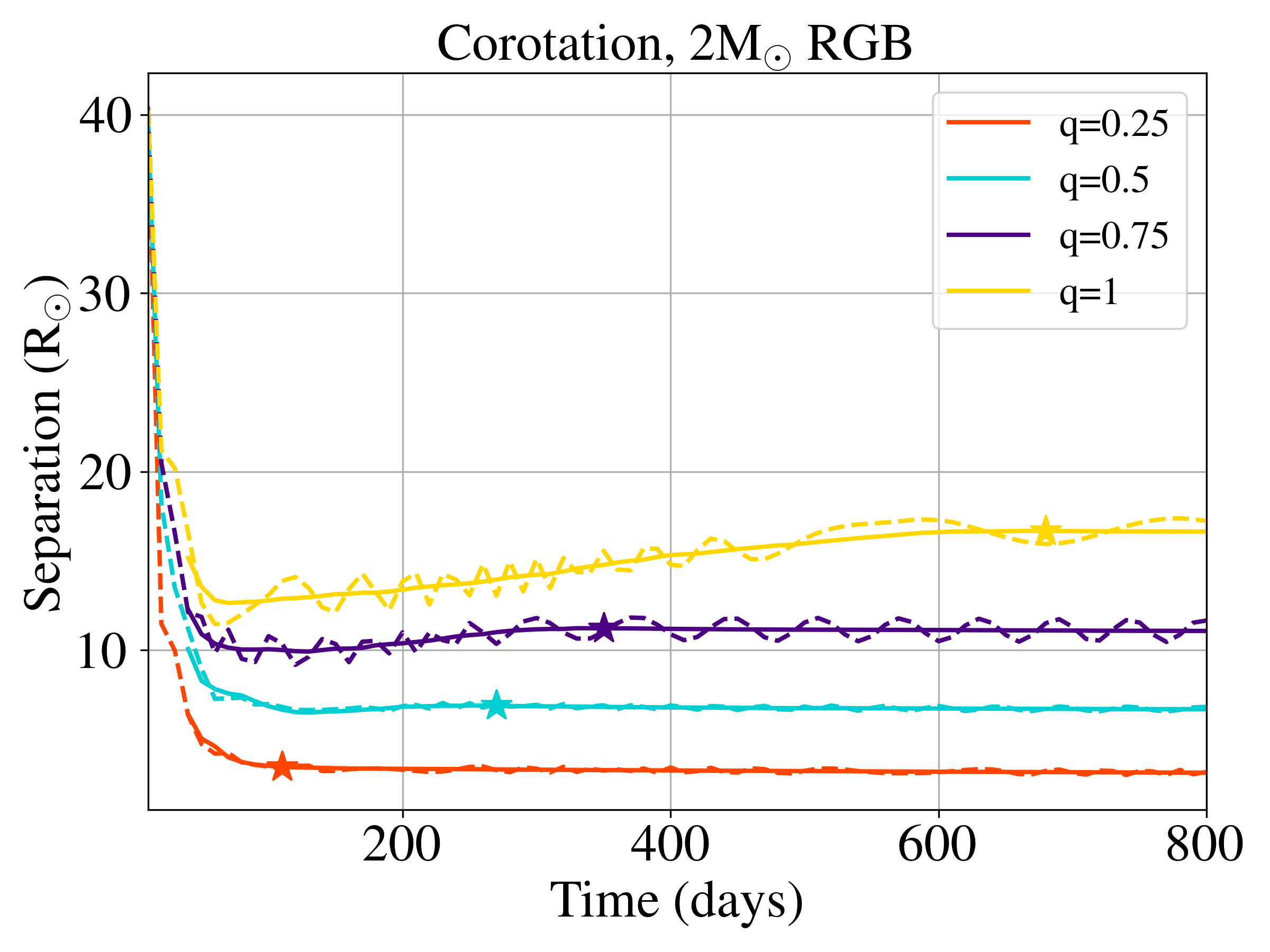}
\includegraphics[width=0.47\textwidth]{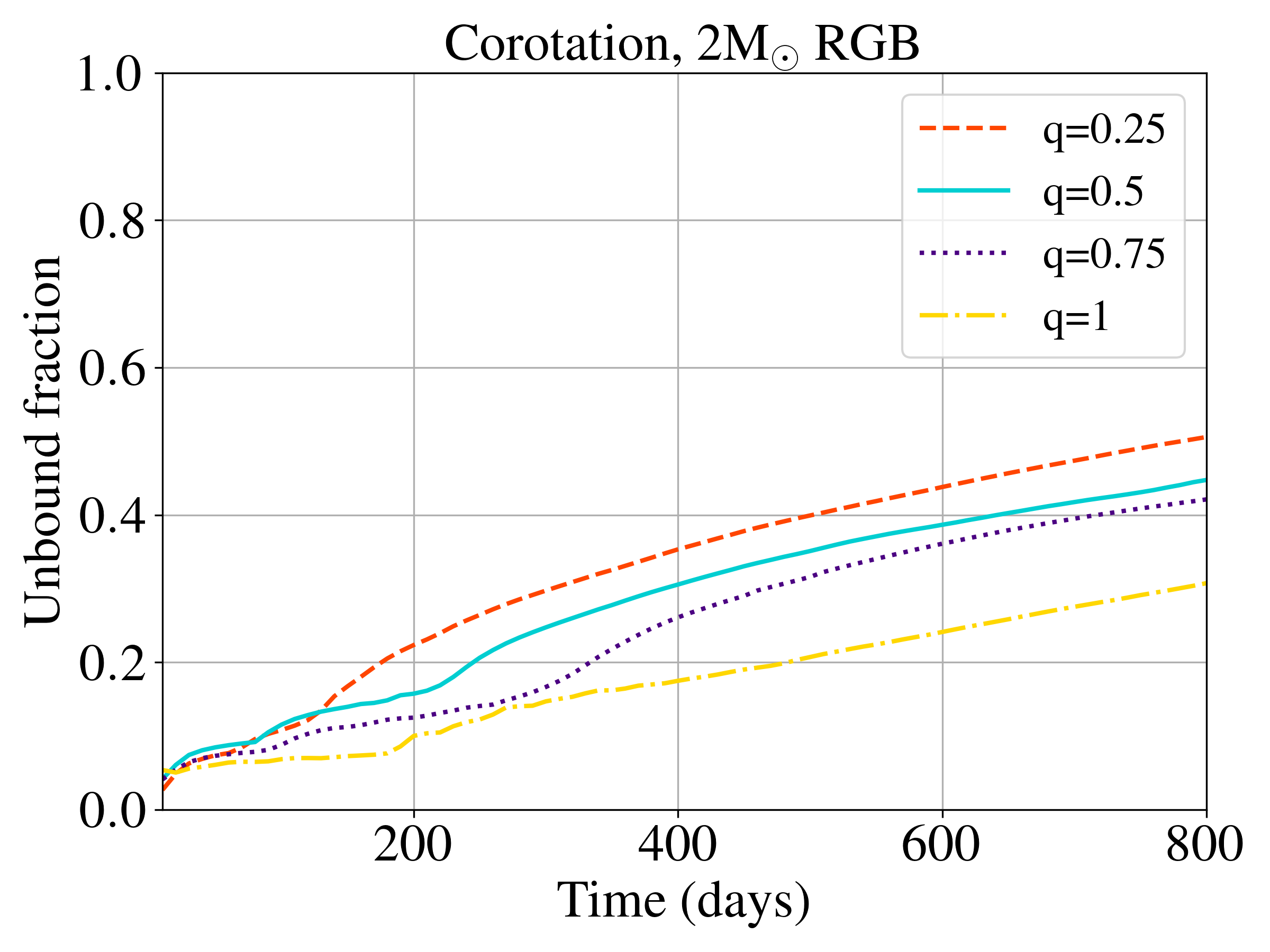}
\includegraphics[width=0.47\textwidth]{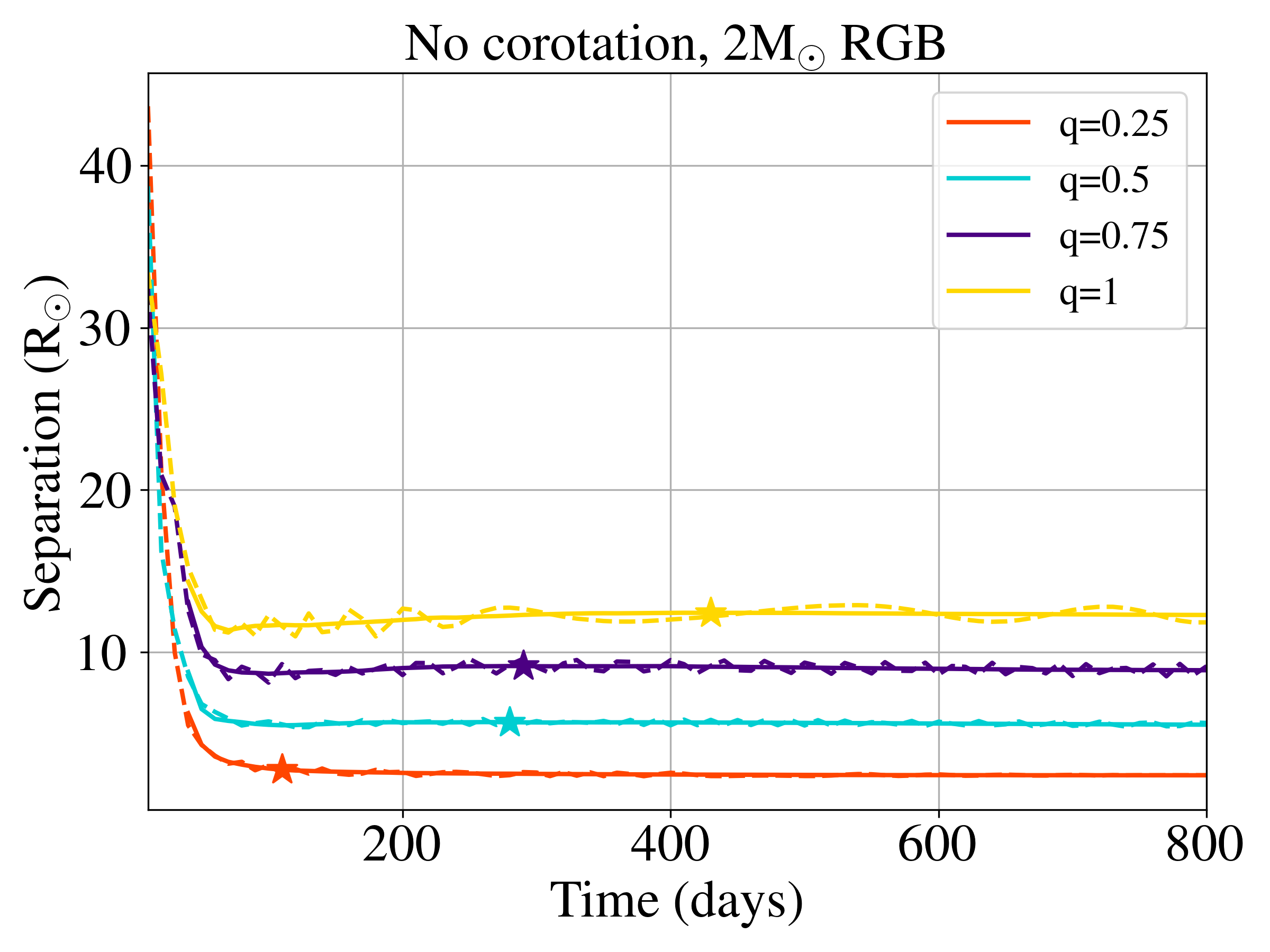}
\includegraphics[width=0.47\textwidth]{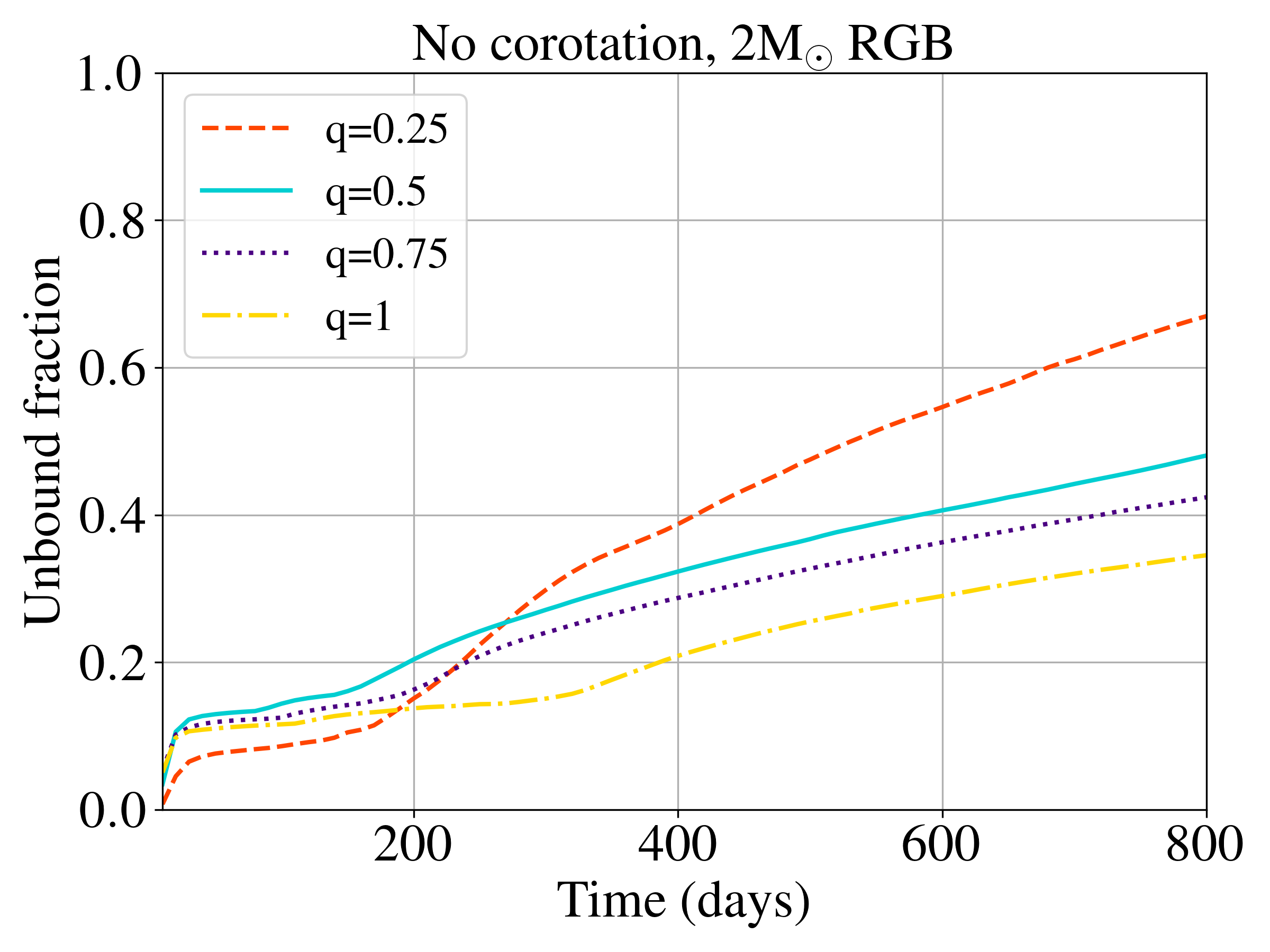}

\caption{
{\it Left:} Change of orbital separation between the core of the giant and the companion over time. The dashed lines represent the instantaneous orbital separation $r$, while the solid lines show the semi-major axis $a$, as defined by equation~\ref{eq:semi-major}. Since the orbit is initially open, $a$ is not shown for the first $\sim 50$ days. The stars show the end of the plunge-in, as defined in the text. \added{Because snapshots from simulations are saved only every 10 days, which is longer than the orbital period, the apparent periodicity in the separation is an sampling artifact rather than the true orbital period}. {\it Right:} Unbound fraction over time. \explain{We included the explanation about the sampling.} }\label{fig:sep_unb}
\end{figure*}

In Fig.~\ref{fig:sep_unb}, we show the instantaneous separation ($r$) between the companion and the core as a function of time, as well as the semi-major axis~($a$), given by:

\begin{equation} \label{eq:semi-major}
a = \left( \frac{2}{r} - \frac{v^{2}}{G M} \right)^{-1},
\end{equation}

\noindent where $v$ is the relative speed between the core and the companion, and $M = M_{\text{1,c}} + M_{\text{2}}$. At early times, the system is in the plunge-in phase of CEE, characterized by a rapid decrease in orbital separation. During the plunge, the \added{relative} orbit \added{between the core and companion} can have $a<0$. After that, the system transitions into the spiral-in phase in which $a$ remains roughly constant.  In Table~\ref{tab:our_simulations}, we list the final separations~(semi-major axis) for the end of the plunge-in~($a_{\rm f}$) for our simulations. For the cases in which $a$ increases before decreasing, we considered the end of the plunge the moment right after it stops increasing. For the cases in which $a$ never increases, we consider the moment in which  $|\dot{a}P/a|=0.001$.

For the $1~\text{M}_\odot$ case, the value of $a$ stalls after end of the plunge phase and no more decay is seen. For the $2~\text{M}_\odot$ case, the separation still slowly decreases, but at a significantly slower rate. For instance, for the corotation cases, the decay at the first year after the end of the plunge phase is $\sim$ 0.23, 0.17, and 0.12 $\text{R}_\odot$ for q=0.25, 0.5, and 0.75, respectively. Fig.~\ref{fig:sep_unb} shows the linear fit between the mass ratio~($q$) and the fitted $a_{\rm f}$~($a_{\rm f,fit}$) normalized by the giant radius~($R_1$) for each set of primary masses. The best linear fit for the $1~\text{M}_\odot$ set (with corotation) is:

\begin{equation}
    \frac{a_{\rm f,fit}}{R_1} = (0.29\pm 0.02) q + (0.02\pm 0.02), \,\, \text{R}^2 = 0.99,
\end{equation}

\noindent for the $2~\text{M}_\odot$ (with corotation) set is:

\begin{equation}
    \frac{a_{\rm f,fit}}{R_1} = (0.33\pm 0.02) q + (0.03\pm 0.03), \,\, \text{R}^2 = 0.99,
\end{equation}

\noindent and for the $2~\text{M}_\odot$ (no corotation) set is:

\begin{equation}
\frac{a_{\rm f,fit}}{R_1}= (0.24\pm 0.01) q - (0.01\pm 0.01), \,\, \text{R}^2 = 0.99,
\end{equation}

\noindent where $\text{R}^2$ is the coefficient of determination.

These correlations can be used to estimate the final separation at the end of the plunge-in phase for systems with mass ratios not simulated. Interestingly, the slope of the linear fit is similar across different primary masses. However, it differs when corotation is included or not in the $2~\text{M}_\odot$ simulations.
We also calculated the final unbound fraction. For that, we considered the sum of mechanical~($E_{\rm mech}$) and internal~($E_{\rm int}$) energies for each particle $i$:

\begin{equation}
   E_{\rm tot} = E_{\rm mech} + E_{\rm int} = m\left(\frac{1}{2}v^2+\phi + Ie\right),
\end{equation}

\noindent where $m$ is the mass of the particle, $v$ is the velocity relative to the center of mass of the bound material, $\phi$ is the gravitational potential, and $Ie$ is the specific internal energy. The relative velocity with respect to the center of mass must be calculated iteratively, since it should only consider the bound particles~\citep[for a thorough discussion, see][]{2019MNRAS.486.5809P}. If $E_{\rm tot,i} < 0$, the particle is bound to the system, while those with $E_{\rm tot,i} > 0$ are unbound. The unbound mass fraction is finally defined as the fractional mass of the envelope with positive total energy. After 800 days, all envelopes are still partially bound.

\subsubsection{Impact of corotation}  \label{sec:corotation}

\begin{figure}
\centering
\includegraphics[width=0.47\textwidth]{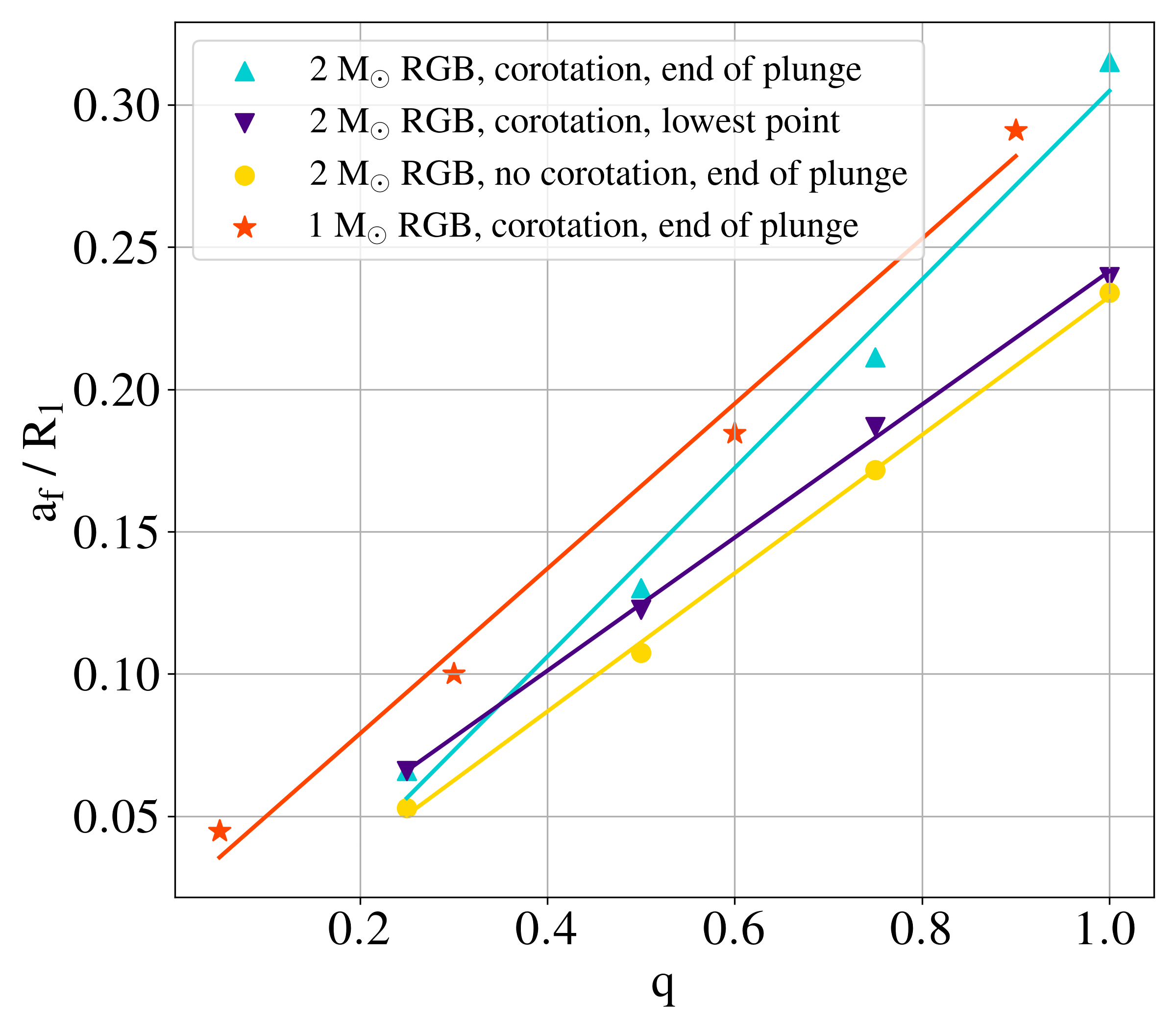}
\caption{Best linear fits for the simulation sets listed in Tab.~\ref{tab:our_simulations}. For the $2~\text{M}_\odot$ case with corotation, two curves are shown: the separation at its minimum (purple), and the separation when the plunge ends, as defined in the text (turquoise).} \label{fig:fit_mysim}
\end{figure}

\begin{figure*}
    \centering
 \includegraphics[width=0.95\textwidth]{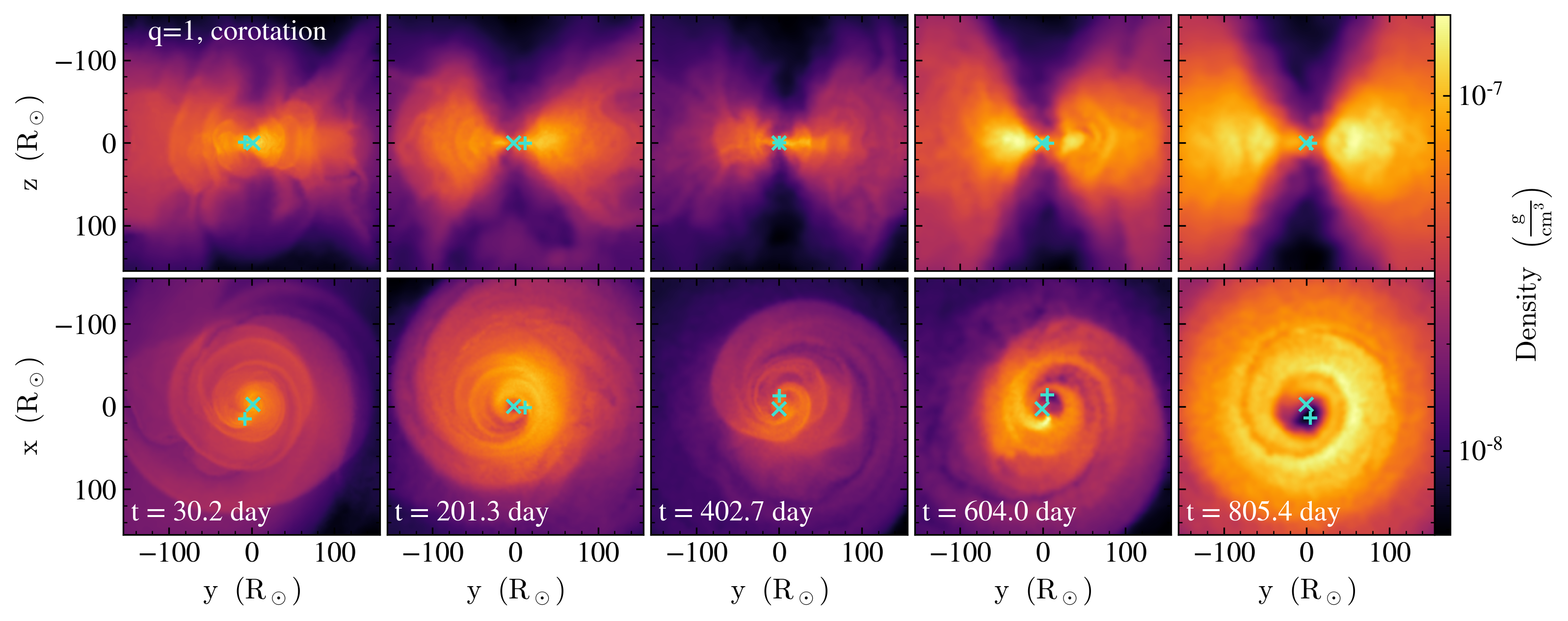}
    \caption{Density in the $y$–$x$ \added{and $y$–$z$} planes at $30.2$, $201.3$, $402.7$, $604$ and $704.7$ days for the $2~\text{M}_\odot$ RGB star, $q=1$ and corotation. The orbit around the companion and the core is mostly cleared by $\sim 700$ days, coinciding with the orbital separation stop increasing. \added{The cyan 'x' is the companion and the '+' is the core.} }
    \label{fig:corotation}
\end{figure*}

To illustrate the effect of corotation, we highlight three curves in Fig.~\ref{fig:fit_mysim}: the non-corotation case at the end of the plunge, and the corotation case both at its minimum and at the end of the plunge. The minimum in the corotation case is very similar to the end of the plunge in the non-corotation run, with the latter being only about 10\% smaller.

The main difference is that, except for $q=0.25$, the corotation cases do not remain at this minimum. After the separation reaches its smallest value, it increases again before leveling off or resuming its decay. This rebound becomes stronger for higher mass ratios. For example, for $q=1$, the separation continues to increase until $\sim$700 days, after which the orbit resumes its decay. The increase in the corotation case occurs because the surrounding material initially carries significant angular momentum and exerts a torque on the point particles, an effect that is much weaker in the non-corotation runs. The similarity between the minimum in the corotation runs and the end of the plunge in the non-corotation case indicates that the rebound is caused by the additional torque exerted by the corotating envelope. Once a cavity around the particles is cleared, as shown in Fig.~\ref{fig:corotation}, no nearby material remains to torque the binary, and the separation stops increasing.

\subsubsection{Impact of the Riemann Solver} \label{sec:riemann}

Our final orbital separation for the $2\,\text{M}_\odot$ system with a mass ratio of $q = 0.5$ is $6.9\,\text{R}\odot$, which is higher than the values reported in the previous \texttt{MANGA} studies~\citep{2019MNRAS.486.5809P, 2023MNRAS.526.5365V}. Some key differences exist between our setup and earlier works, as detailed in Section~\ref{sec:methods}, the most significant being the number of particles and the choice of the Riemann solver. To investigate this discrepancy, we performed an additional simulation using the more diffusive \texttt{HLL} solver with the same number of particles as our other simulation (1.2 million star particles, 2 million total). This yielded a final separation of $3.8\,\text{R}\odot$ after the rapid plunge, close to the $3.6\,\text{R}\odot$ reported by \citet{2019MNRAS.486.5809P} for a simulation with 320,000 particles. This separation is higher compared to the $\sim 5\,\text{R}\odot$ found by \citet{2023MNRAS.526.5365V}, which is probably due to their lower resolution (80,000 particles).

Thus, we conclude that the difference in the separation between this paper and previous studies is mostly due to the Riemann solver. Since the \texttt{HLLC} solver is less diffusive, we argue that our results in this paper are robust. Furthermore, our final orbital separation lies between the values reported by \citet{2016ApJ...816L...9O} and \citet{2018MNRAS.480.1898C}, who found separations of $5\,\text{R}\odot$ and $8\,\text{R}\odot$, respectively, for a similar system. \added{Although differences in simulation duration, numerics, and physics prevent a strictly direct comparison, the fact that our $6.5\,\text{R}\odot$ result falls within this range suggests that our results are broadly consistent with what is expected from previous studies.}

\section{Comparison with previous simulations}
\label{sec:other_simulations}

\begin{figure*}
\centering

 \includegraphics[width=0.47\textwidth]{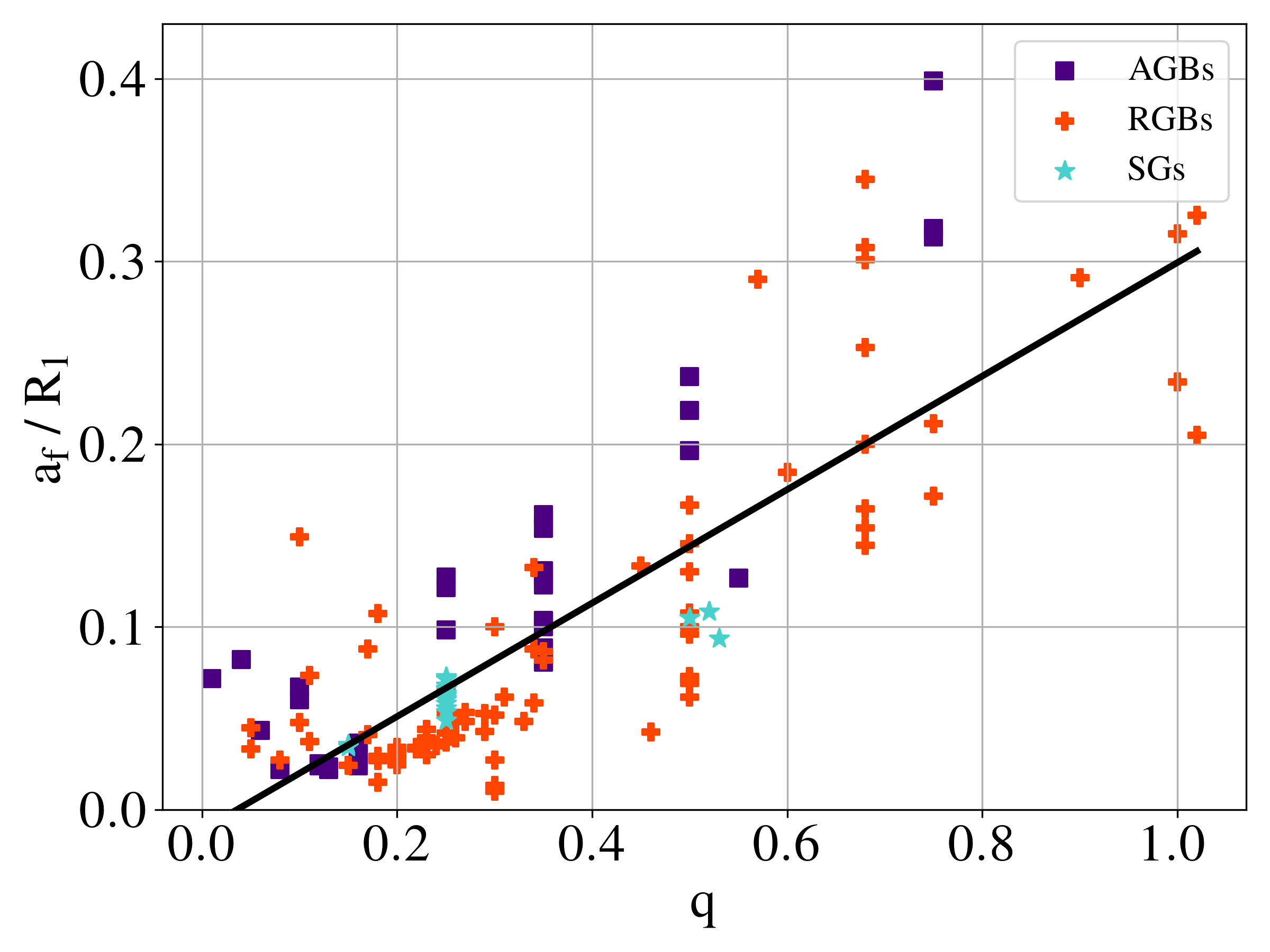}
  \includegraphics[width=0.47\textwidth]{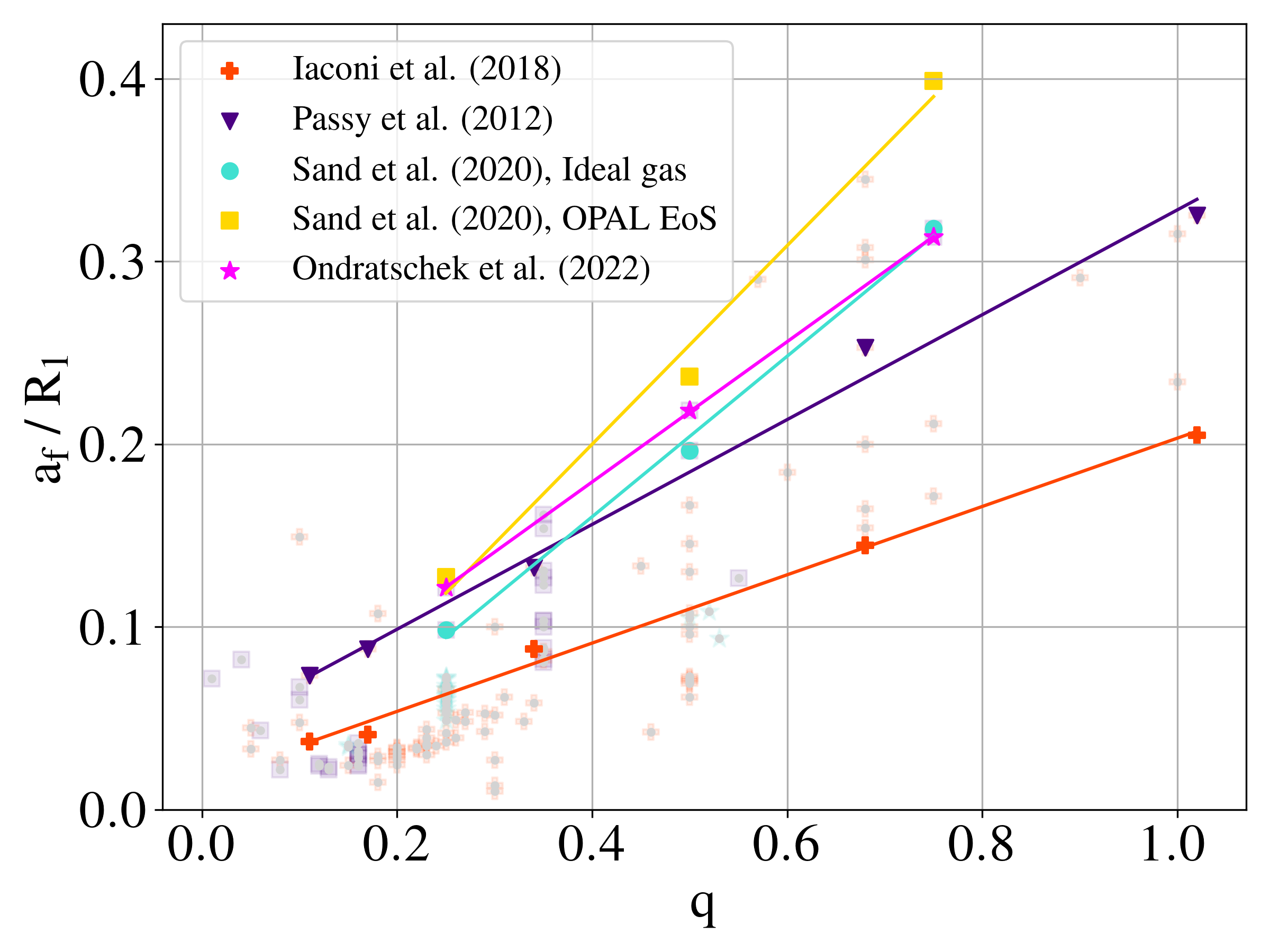}
\caption{ {\it Left}: AGB, RGB, and Supergiants - SGs (Tabs.~\ref{tab:A3} and~\ref{tab:new_sims}) simulations from the literature, along with the best collective fit (black). {\it Right}: Comparison of the linear fit for different sets of simulations from the literature that studied different q keeping other parameters fixed~\citep{2012ApJ...744...52P,2018MNRAS.477.2349I,2020A&A...644A..60S,2022A&A...660L...8O}. The shadowed colored markers represent all simulated systems.}

\label{fig:all_correlation}
\end{figure*}

In this Section, we expanded our analysis to include additional simulations from the literature. In Fig.~\ref{fig:all_correlation}.a and Tab.~\ref{tab:correlations}, we present the simulation sets and the best fits for \citet{2012ApJ...744...52P}, \citet{2018MNRAS.477.2349I}, \citet{2020A&A...644A..60S}, and \citet{2022A&A...660L...8O}. In each of these studies, the primary mass and radius are kept fixed while the mass ratio is varied. 

An important case is the AGB simulation sets from \citet{2020A&A...644A..60S} and \citet{2022A&A...660L...8O}, which use a significantly more evolved primary with a higher core mass ($ M_{\rm 1,c} = 0.545 \, \text{M}_\odot $), in contrast to the RGB models used in this work, as well as in \citet{2012ApJ...744...52P} and \citet{2018MNRAS.477.2349I} ($ M_{\rm 1,c} = 0.37{-}0.39 \, \text{M}_\odot $). Despite these differences in core mass, all systems agree with a linear correlation.

\begin{table}
    \centering
    \caption{Best linear fits in the form $a x + b$ for the simulation sets shown in Fig.~\ref{fig:all_correlation}a. The masses are in M$_\odot$ and distances in R$_\odot$.}
    \label{tab:correlations}
    \begin{tabular}{c c c c c c c}
        \hline\hline
        $M_{1,\mathrm{t}}$ & $M_{1,\mathrm{c}}$ & $R_1$ & $a$ & $b$ & R$^2$ & Ref. \\
        \hline
        0.88 & 0.39  & 83  & 0.19 &  0.016 & 0.99   & (1) \\
        0.88 & 0.39  & 83  & 0.28 &  0.042 & 0.99 & (2)$^\dagger$ \\
        0.97 & 0.545 & 173 & 0.44 &  -0.015 & 1   & (3)$^*$ \\
        0.97 & 0.545 & 173 & 0.54 &  -0.017 & 0.99  & (3)$^+$ \\

        0.97 & 0.545 & 173 & 0.38 &  0.026  & 0.99   & (4) \\
        \hline
    \end{tabular}
    \vspace{0.2cm}
    
    $\dagger$:SIM1-SIM5; $^*$: ideal gas;  $^+$: OPAL EoS.
    \textbf{References:}  (1)~\citet{2018MNRAS.477.2349I}; 
    (2)~\citet{2012ApJ...744...52P};(3)~\citet{2020A&A...644A..60S}; (4)~\citet{2022A&A...660L...8O}
\end{table}

Another point of interest is that \citet{2018MNRAS.477.2349I} used an identical primary and the same code as \citet{2012ApJ...744...52P}, but a different gravity solver. Despite the shared physical setup, the slope of \citet{2012ApJ...744...52P} is closer to our results than that of \citet{2018MNRAS.477.2349I}. This shows that changes caused by numerics can be greater than those caused by different masses or evolutionary stage. This agrees with our discussion in Sec.~\ref{sec:riemann}, in which we conclude that the Riemann solver can have substantial impact on the final separation. For this reason, even though we expect some variation for different parameters, e.g. AGB system from \cite{2020A&A...644A..60S} and \cite{2022A&A...660L...8O} seems to have a higher slope than the RGBs from this work, \cite{2012ApJ...744...52P} and \cite{2018MNRAS.477.2349I}, it remains challenging to isolate how the slope of the fit changes across different giant parameters, since the impact of numerics may be comparable than the one from different physical setups. To guarantee consistency, any comparison of different physical parameters should be done with the same code, solvers and resolution.

Since these simulation sets yield similar linear fits, we also sought a general expression that includes all simulations. We began with Tab.~A3~(see Tab.~\ref{tab:A3}) from \citet{2019MNRAS.490.2550I}, and added subsequent published simulations with supergiants, RGBs and, AGBs~(Tab.~\ref{tab:new_sims}) as primaries. Although our compilation contains the majority of simulations, it is not exhaustive.

We excluded the supergiant simulations of \citet{2020arXiv201106630L} because their models start with a much smaller initial separation, placing the secondary deep within the envelope from the outset. Consequently, the resulting values of $a_{\rm f}$ are substantially lower than those of other studies. Similarly, we exclude \citet{2021MNRAS.507.2659G} because their simulations \added{have non-zero initial eccentricity, and are thus difficult to compare with the others, which have zero initial eccentricity. We have also excluded the work from \citet{2025ApJ...979...57L} and one simulation from \citet{2022MNRAS.512.5462L}, because their orbital separation reaches a value equal to the softening length before the end of simulations}. Finally, within Tab.~\ref{tab:new_sims}, we include only simulations that provide unique combinations of parameters and outcomes. Specifically, if a given paper \added{(or group of papers using the same code)} presents multiple simulations with similar initial parameters and identical final separations, we treat them as a single entry. In contrast, if they have different final separations even if the initial conditions are the same, each is included as a separate entry.

\added{One source of concern when getting $a_f$ from other studies was whether to use the separation at the end of the plunge-in phase or the separation at the end of the simulation, in cases where both were reported \citep[e.g.][]{2020A&A...644A..60S, 2020MNRAS.494.5333R, 2022MNRAS.517.3181G}. We opted to adopt the separation at the end of the simulations of each study, even though our main thesis is that the linear relation holds for the separation at the end of the plunge. The reason is that in all those studies that provide both values, the end of the plunge is defined using a different criterion than ours ($|\dot{a}P/a| = 0.01$, compared to our $0.001$). This difference in definition alone can account for some of the additional orbital decay they report after their nominal end of the plunge. For illustration, in our own $q=0.25$, $2\,\text{M}_\odot$ RGB simulations (with corotation), the end-of-plunge separation is $6.4\,\text{R}_\odot$, $4.1\,\text{R}_\odot$, or $3.5\,\text{R}_\odot$ for thresholds of $|\dot{a}P/a|=0.1$, $0.01$, and $0.001$, respectively. This can also be seen in Fig.\ 6 of \citet{2022MNRAS.517.3181G}, where their separation seems to reach a plateau, but their criteria sometimes marks the end-of-plunge before this. 

This shows that the definition for the end of the plunge can also have an impact on the slope. For this reason, we emphasize that our definition was motivated by physical considerations: we required that the core+companion system had just reached the quasi-plateau in separation and that the envelope was close to reorganizing into its new morphology (see Fig.~\ref{fig:corotation} for the morphology evolution). We found that a threshold of $|\dot{a}P/a|=0.001$ better satisfies these two criteria than $0.01$, and therefore adopted this value. One key issue with considering a value of 0.01 is that it leaves a small gap between the plunge and the slow spiral-in~\citep[see, e.g., Fig.~1 of][]{2016MNRAS.462..362I}. In this work, we chose a definition that effectively extends the plunge, so there is no such gap. In this sense, our $a_f$ can be equivalently referred to as either the separation at the beginning of the spiral-in or the post-plunge separation.

A final complication is that some simulations may have been stopped before the plunge was truly complete. To avoid cherry-picking, we kept all simulations that this may have happened, while noting that this contributes with additional scatter to the relation. Taken together with the discussion above, this reinforces that although previous simulations broadly follow a linear correlation, future work aiming to refine this empirical relation must consider not only consistent numerics and physics, but also a standardized definition for the end of the plunge and enough time to guarantee that the plunge can fully be completed.}

Fig.~\ref{fig:all_correlation}.b shows all simulations presented in Tabs.~\ref{tab:A3} and~\ref{tab:new_sims}, along with the best linear fit:

\begin{equation}\label{eq:all_correlation}
    \frac{a_{\rm f, fit}}{R_1} = (0.31 \pm 0.02)\, q - (0.011 \pm 0.007), \quad \text{R}^2 = 0.68.
\end{equation}

This result shows that the final separation at the end of the plunge-in phase can be reasonably predicted from $q$ using a linear fit. Although the precise fit may vary depending on the type of giant star, the linear correlation remains a useful approximation, especially since numerics often introduce the largest uncertainties. 

An equivalent formulation using the initial separation gives comparable results:
\begin{equation}\label{eq:all_correlation_ai}
    \frac{a_{\rm f, fit}}{a_i} = (0.26 \pm 0.01)\, q - (0.017 \pm 0.006), \quad \text{R}^2 = 0.69.
\end{equation}

\explain{The tiny difference in the fit from submission 2 comes from the change of $a_f$ from \cite{2020MNRAS.494.5333R} to be from the end of simulations and not end of plunge, and the exclusion of simulations from \citet{2025ApJ...979...57L}. }

Even though the $a_i$-normalized fit shows marginally better correlation ($\text{R}^2 = 0.69$ versus $0.68$), we consistently use ${a_{\rm f, fit}}/{R_1}$ throughout this work for simplicity, since $R_1$ is the original parameter reconstructed from observations, whereas $a_i$ would need to be further estimated from $R_1$ and $q$. 

\begin{figure}
\centering
 \includegraphics[width=0.47\textwidth]{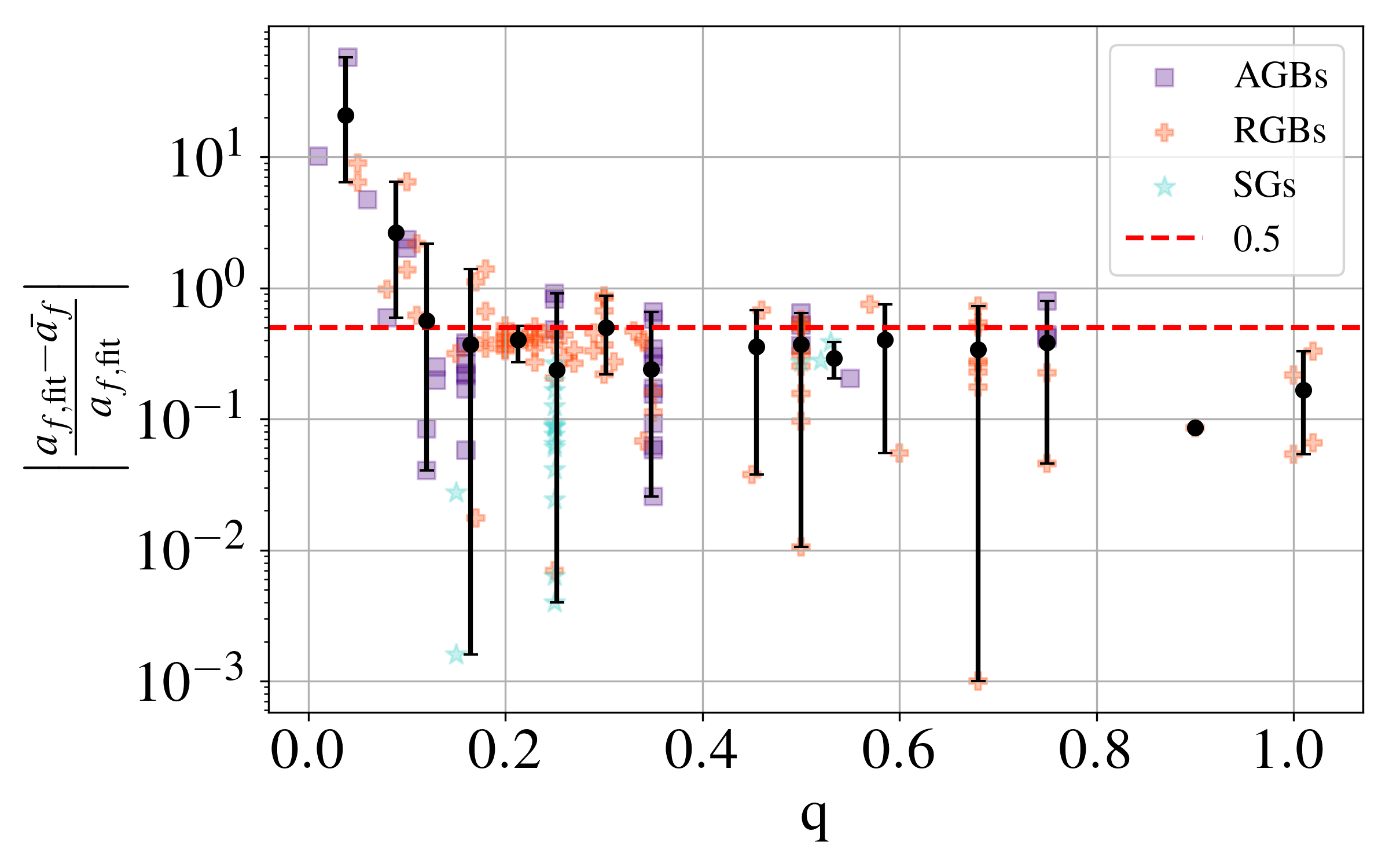}
\caption{
Simulations are grouped into 22 equally spaced q-bins. Each point represents the mean absolute deviation normalized by the fitted value, with error bars showing the full range (min to max) within each bin, not the standard deviation. The shadowed colored markers represent all simulated systems.
}\label{fig:residuals}
\end{figure}

We also assess the reliability range of the fit. As a threshold, we assume that numerical differences may cause variations up to half of $a_{\rm f, fit}$. We compute the relative difference between $a_{\rm f}$ and $a_{\rm f, fit}$, normalized by $a_{\rm f, fit}$, and show the results in Figure~\ref{fig:residuals}. This difference remains below 0.5 for most of the range, but exceeds it for $q \lesssim 0.15$, suggesting that the fit becomes unreliable there.

\section{Comparison with the energy formalism}
\label{sec:formalism}
\begin{figure*} \centering 
\includegraphics[width=0.45\textwidth]{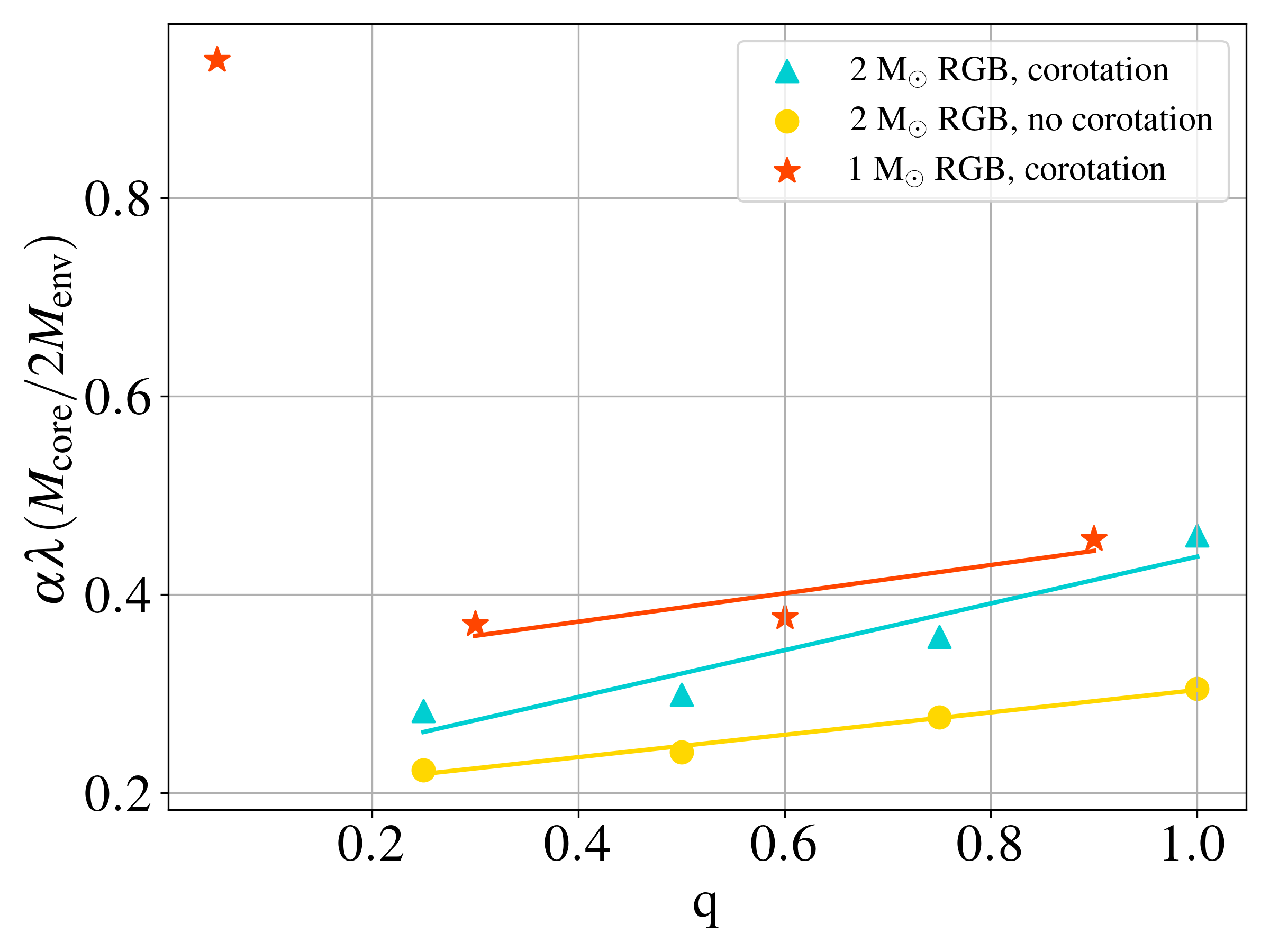} \includegraphics[width=0.45\textwidth]{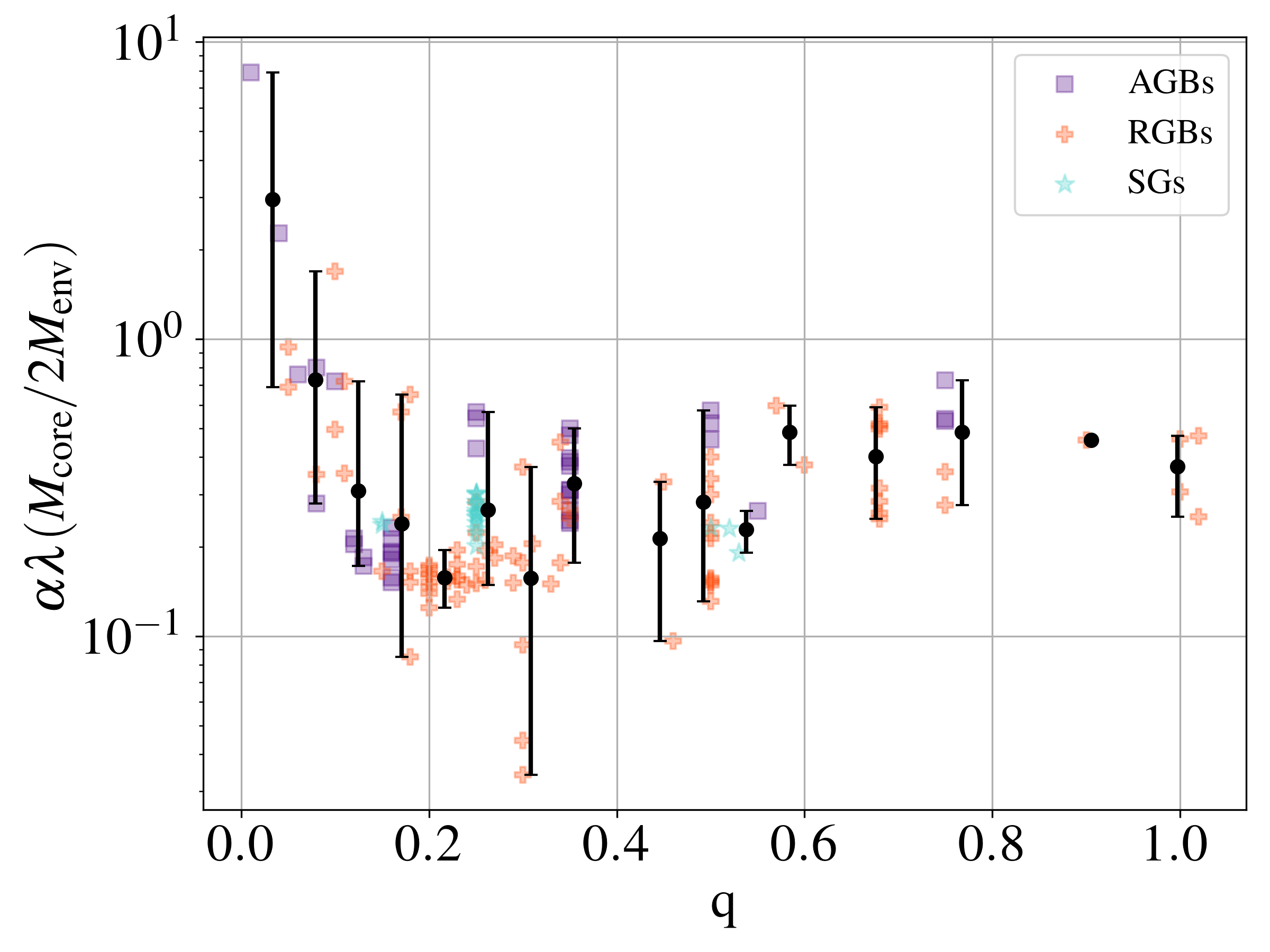}

\caption{{\it Left:} Best linear fits between the mass ratio $q$ and $\alpha \lambda /(M_{1,\rm c}/2M_{1,\rm e})$ for our simulations sets. The $q=0.05$, $1\,\text{M}_\odot$ point is plotted but excluded to the fit due to its departure from the overall trend. {\it Right:} Plot of q versus $\alpha \lambda /(M_{1,\rm c}/2M_{1,\rm e})$ for all simulations. Following the same procedure of Fig.~\ref{fig:residuals}, simulations are grouped into 22 equally spaced $q$-bins and each black point shows the mean value, while the vertical error bars indicate the full range of values within each bin. The shadowed colored markers represent all simulated systems.}
 \label{fig:alpha} \end{figure*}
Mathematically, the energy formalism can be expressed as:

\begin{equation} \label{eq:alpha} E_{\text{env}} = \alpha \Delta E_{\text{orb}} , \end{equation}

\noindent where $E_{\text{env}}$ represents the energy required to unbind the envelope of the giant, and $\Delta E_{\text{orb}}$ is the change in orbital energy. Unfortunately, choosing specific expressions for $E_{\text{env}}$ and $\Delta E_{\text{orb}}$ is not straightforward. Arguably, the most widespread version~\citep{1984ApJ...277..355W,1990ApJ...358..189D} of the formalism is:

\begin{equation} \label{eq:webbink} G\frac{M_{1,\text{t}}M_{1,\text{e}}}{\lambda R_1} = \alpha\Bigg[\frac{GM_{1,\text{c}}M_2}{2a_{\rm f}}-\frac{GM_{1,\text{t}}M_2}{2 a_{\rm i}}\Bigg]. \end{equation}

\noindent where $a_{\rm i}$ is the initial orbital separation, $M_{1,\text{e}} =  M_{1,\text{t}} - M_{1,\text{c}}$ is the envelope mass and $\lambda$ is a parameter related to the structure of the envelope. The value of $\lambda$ is often tabulated for different stellar masses, evolutionary stages, and the contribution of internal energy~\citep[see, e.g.,][]{ 2010ApJ...716..114X}. 

Alternative equations for the formalism have also been proposed~\citep[e.g.,][]{1993PASP..105.1373I, 2002MNRAS.329..897H, 2011MNRAS.411.2277D}. Unfortunately, this means that the definitions of $\alpha$ and $\lambda$ are not standardized. Another possible formalism is~\citep{2002MNRAS.329..897H,2006MNRAS.369.1152K}:

\begin{equation} \label{eq:hurley} G\frac{M_{1,\text{t}}M_{1,\text{e}}}{\lambda R_1} = \alpha\Bigg[\frac{GM_{1,\text{c}}M_2}{2a_{\rm f}}-\frac{GM_{1,\text{c}}M_2}{2 a_{\rm i}}\Bigg], \end{equation}

\noindent in which the only difference from Eq.~(\ref{eq:webbink}) is the use of $M_{1,\text{c}}$ instead of $M_{1,\text{t}}$ to estimate the initial orbital energy.

\subsection{Does the energy formalism expect a linear fit?}

To understand the conditions under which a linear relation could emerge from the energy formalism, we begin by rearranging Eq.~(\ref{eq:webbink}) as:

\begin{equation}
\frac{2a_{\rm i}}{\alpha \lambda} \frac{ M_{1,\text{e}}}{ M_2 R_1} = \frac{M_{1,\text{c}}a_{\rm i}}{M_{1,\text{t}}a_{\rm f}} - 1.
\end{equation}
If $ \frac{M_{1,\text{t}}a_{\rm f}}{M_{1,\text{c}}a_{\rm i}} < 1$, the expression can be rewritten as:
\begin{equation}
\frac{M_{1,\text{t}}a_{\rm f}}{M_{1,\text{c}}a_{\rm i}} + \left( \frac{M_{1,\text{t}}a_{\rm f}}{M_{1,\text{c}}a_{\rm i}} \right)^2 + \mathcal{O}(3) = \left( \alpha \lambda \frac{R_1}{a_{\rm i}} \right) \frac{M_{1,\text{t}} q}{2 M_{1,\text{e}}}.
\end{equation}

If $ \frac{M_{1,\text{t}}a_{\rm f}}{M_{1,\text{c}}a_{\rm i}} \ll 1$, higher-order terms become negligible, and the expression simplifies to a linear form:
\begin{equation} \label{eq:linear_webbink}  
\frac{a_{\rm f}}{R_1} \simeq (\alpha \lambda) \frac{M_{1,\text{c}} q}{2 M_{1,\text{e}}} 
\end{equation}

The same expression can be derived from Eq.~(\ref{eq:hurley}) under the condition that $ a_{\rm f} \ll a_{\rm i}$, which is easier to satisfy since $ M_{1,\text{c}}< M_{1,\text{t}}$. Given that in the limit  $\frac{M_{1,\text{t}}a_{\rm f}}{M_{1,\text{c}}a_{\rm i}} \ll 1$, both equation converge, it is not surprising that \citet{2010A&A...520A..86Z} found similar $\alpha$ from observations by considering Eqs.~\ref{eq:webbink} and \ref{eq:hurley}.

Interestingly, many simulations (including those presented here) do not completely unbind the envelope, a prerequisite for using the energy formalism, yet still exhibit a relation resembling Eq.~\ref{eq:linear_webbink}. Previous studies have noted that drag forces will cease once the local envelope density falls below a critical threshold, when co-rotation is reached, or when orbital motion becomes subsonic~\citep{2023LRCA....9....2R}, rather than when sufficient energy has been released to unbind the envelope. This decoupling, also discussed by \citet{2022PhRvD.106d3014T}, implies that the in-spiral can stall before or after the envelope has enough energy to be ejected. Hence, the observed linearity between $a_{\rm f}/R_1$ and $q$ likely arises from the dynamical evolution itself, not from energy balance.

As noted by \citet{2019MNRAS.490.2550I}, $\alpha$ derived from simulations may be better interpreted as a proportionality constant rather than a true efficiency. In this view, $\alpha$ quantifies how the envelope binding energy scales with the change in orbital energy at the point where the plunge stalls. Stopping drag requires lifting the envelope, and more tightly bound envelopes are harder to raise, which naturally explains why a relation linking the final separation and the binding energy (such as the energy formalism) produces a linear dependence similar to that seen in simulations. In this interpretation, values of $\alpha>1$ after the plunge, when no additional energy sources are present, simply indicate that insufficient orbital energy has been released, and that the system will therefore continue its decay during the subsequent slow spiral-in phase.

The idea that the post-plunge separation is governed by dynamical friction rather than energy conservation has been widely discussed. For instance, \cite{2021A&A...648L...6P}, building on \cite{1995ApJ...451..308Y} and \cite{1996ApJ...458..692T}, proposed that the plunge ends once the frictional torque on the companion approaches zero. In their approach, the final separation is determined dynamically from the condition of vanishing torque, and only afterward is the released orbital energy compared to the envelope’s binding energy. Although the energy formalism is still invoked to determine whether the envelope was unbound, the separation itself is not determined by $\alpha$. Their model predicts a post-plunge separation smaller than expected from pure energy arguments, whereas simulations generally indicate that the plunge halts before sufficient energy is released. Nonetheless, the underlying conclusion is the same: the final separation should not be derived from energy arguments alone.

\subsection{How the slope $\alpha \lambda M_{1,\rm c}/2M_{1,\rm e}$ changes with $q$?}
\label{sec:alpha}
Comparing equations~\ref{eq:all_correlation} and~\ref{eq:linear_webbink} suggest that $\alpha \lambda M_{1,\rm c}/2M_{1,\rm e}$ should be roughly constant on $q$. In other words, $\alpha$ should scale linearly with $M_{1,\text{e}}$, $1/\lambda$, and $1/M_{1,\text{c}}$, and to depend only weakly on $q$.  To test this, we plot $\alpha \lambda M_{1,\rm c}/2M_{1,\rm e}$ (estimated by rearranging Eq.~\ref{eq:hurley}) as a function of $q$ for our set of simulations and for all simulations (Figs.~\ref{fig:alpha}.a and b, respectively). Across the mass ratios, $\alpha \lambda M_{1,\rm c}/2M_{1,\rm e}$ remains roughly constant for $q>0.15$, with only a small increase for $q\sim 1$ compared to $q\sim 0.2$. For $q<0.15$, the parameter starts to deviate and becomes substantially higher. This weak $q$-dependence (except for $q<0.15$) explains why the linear approximation works well within $0.15<q<1$ but becomes unreliable for $q<0.15$, as discussed in Sec.~\ref{sec:other_simulations}. 

From Fig.~\ref{fig:alpha}a, we can also assess how $\alpha$ varies with $q$ for our simulations. For this set of simulations, $\lambda$, $M_{1,\mathrm{c}}$, and $M_{1,\mathrm{e}}$ are constant for $q$, which means that only $\alpha$ varies. Although the slope of this variation is shallow, the fit of $\alpha$ with $q$ is consistent with a linear fit, except for very low mass ratios (in our case, $q=0.05$), where the simulations deviate significantly from the overall fit. Given that $\alpha$ varies linearly with $q$, Eq.~\ref{eq:linear_webbink} then implies that $a_{\mathrm{f}}/R_1$ should depend quadratically on $q$, rather than linearly. However, because the variation of $\alpha$ is small and the simulations cover only about one order of magnitude in $q$, any quadratic contribution remains small, and the relationship is approximately linear within the sampled range. Nevertheless, extrapolating this linear fit to extreme mass ratios ($q>>1$) is not recommended.

\section{Comparison with observations} \label{sec:observations}

Comparison of simulations with observations is inherently challenging due to the uncertainties and limitations of both approaches. Simulations often exclude key physics, such as radiation, and cover only certain phases of CEE. Observations, meanwhile, must reconstruct pre-CEE parameters from post-CEE systems, a process that involves numerous assumptions. Despite these challenges, such comparisons provide valuable insight into how well simulations reproduce observed phenomena.

\begin{figure*}
\centering
\includegraphics[width=0.47\textwidth]{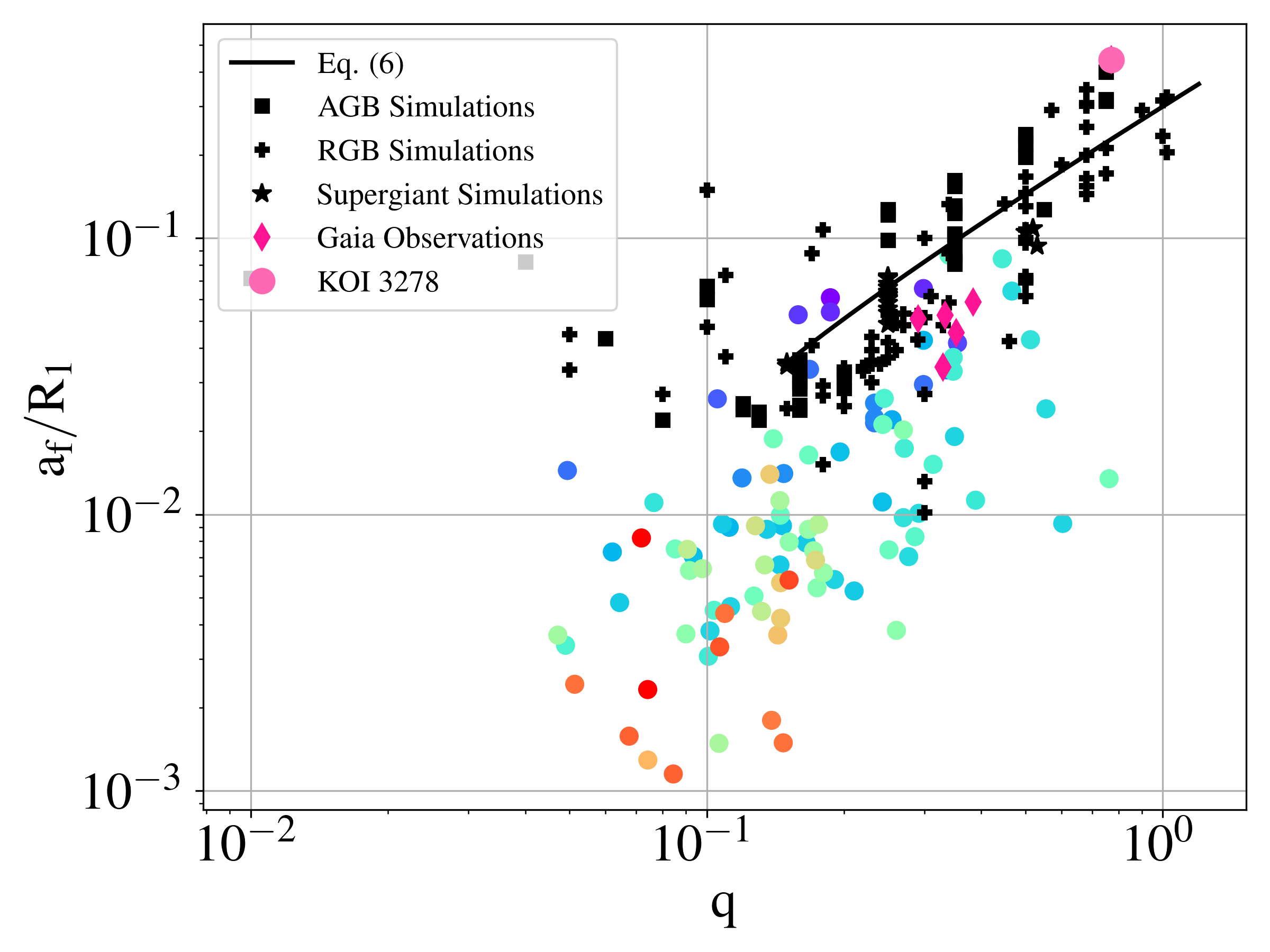}
\includegraphics[width=0.47\textwidth]{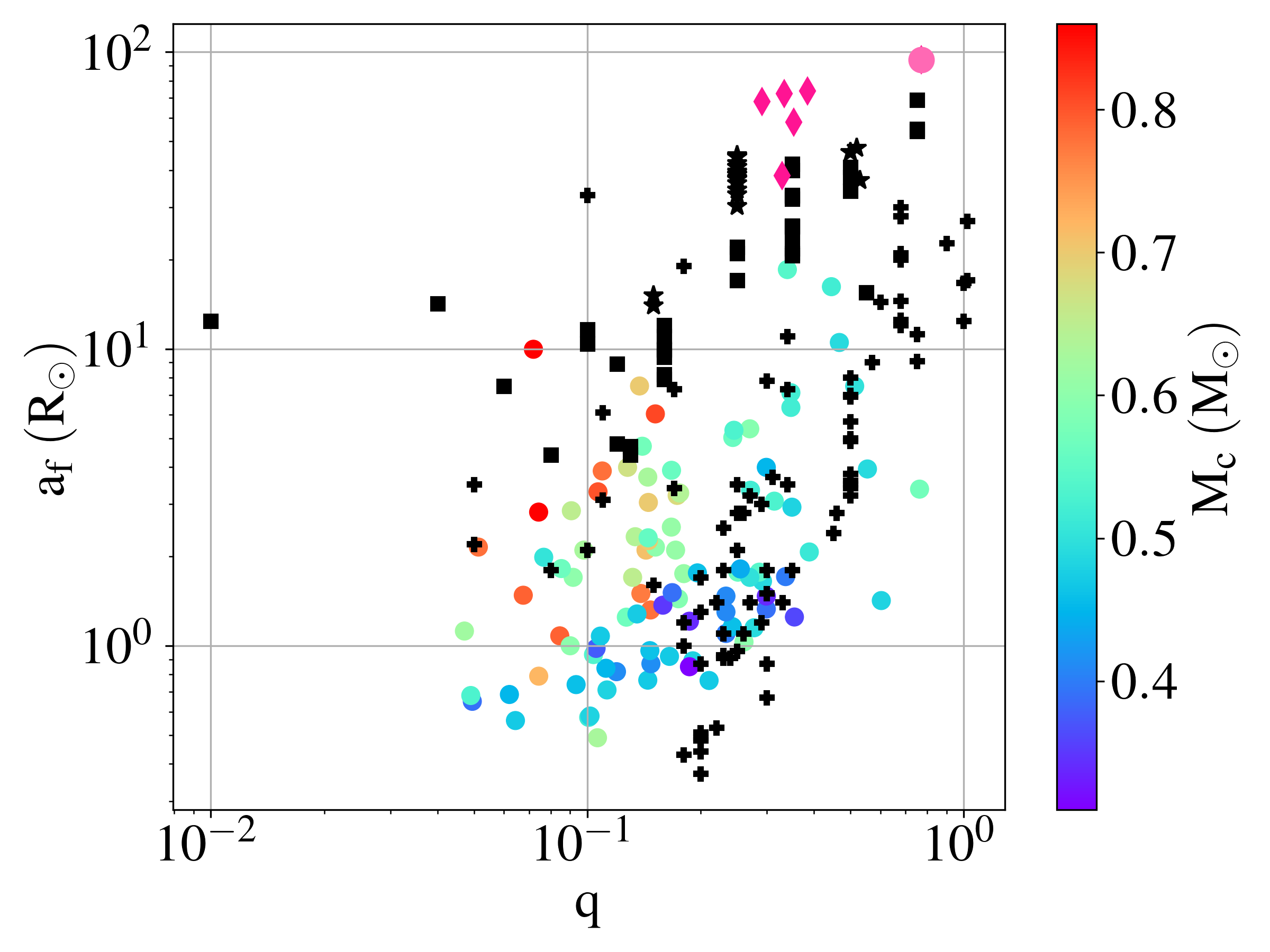}
\caption{{\it Left}: Ratio of the final separation to the giant radius as a function of the mass ratio for observations and simulations (in black), along with the linear fit from Eq.~(\ref{eq:all_correlation},black line); {\it Right}: Final separation as a function of the mass ratio for observations and simulations. The observations are color-coded by the mass of the core, except the intermediate-period systems,  J2117+0332, J1111+5515, J1314+3818, J2034-5037, J0107-2827, unveiled by {\it Gaia}~($M_{\rm1, c}>1~\text{M}_\odot$, pink diamonds), and KOI-3278~($M_{\rm1, c}=0.53~\text{M}_\odot$, pink circle).}
\label{fig:observations}
\end{figure*}

For this analysis, we considered the post-CEE observations listed in \citet{2021ApJ...920...86K}, but did not use the entire dataset. First, we excluded systems flagged as cataclysmic variables, undergoing mass transfer, or identified as triples, as our goal was to focus on binaries whose separations are likely to reflect the immediate post-CEE state. Second, we removed systems with WDs, neutron stars, or black hole secondaries, since these could represent outcomes of a second CEE. Third, we limited our sample to systems in which $M_{\rm1, c}$ is between 0.2 and 0.87~M$\odot$. The lower limit guarantees that the giant radius is at least as large as $a_{\rm f}$ found in our simulations. This estimate is based on a 1~M$_\odot$ star and the $M_{\rm1, c}$–$R$ relation from \citet{1987ApJ...319..180J}. The upper limit corresponds to the maximum core mass for which the reconstruction method described below is applicable. After reconstruction, we also excluded the systems in which $q>1$. In the end, we had 96 systems for comparison~\footnote{Although we do not include a table listing these systems, we provide a Python script that generates Fig.~\ref{fig:observations} from the original machine-readable table published by \citet{2021ApJ...920...86K}.}.

To reconstruct the pre-CEE parameters, we followed the procedure described in \cite{2019MNRAS.490.2550I}, which is based on the method from \cite{2011MNRAS.411.2277D}. For completeness, we briefly outline the steps below (for a detailed discussion, please refer to their papers). The first step is to determine whether the pre-CEE giant was on the AGB or RGB branch. There is no definitive method, but the authors use the following rule. If $ M_{\rm1, c} \leq 0.47~\text{M}_\odot$, the star went through CEE during the RGB phase; if $ M_{\rm1, c} > 0.47~\text{M}_\odot$, it was in the AGB phase.

For post-RGB stars, the authors adopted an initial main-sequence mass of $M_{\rm MS} = 1.19~\text{M}_\odot$. Although this may appear to be a rough approximation, it is justified by the fact that low-mass giants ($M_{1,\text{t}} \sim 1~\text{M}_\odot$) reach similar maximum radii at the tip of both the RGB and AGB phases. In contrast, more massive stars ($M_{1,\text{t}} \gtrsim 2~\text{M}_\odot$) reach significantly larger radii during the AGB phase~\citep[see Fig 4.1 of ][]{2019ibfe.book.....B}. Since Roche lobe overflow is a necessary precondition for CEE, post-RGB systems are more likely to originate from low-mass stars.

For AGBs, the authors provide a figure that relates the WD mass~($M_{\rm WD}$) to the original main-sequence mass~(up to $M_{\rm MS}=5$~M$_{\odot}$). To account for the interruption in the star's evolution, the authors increase $M_{\rm1, c}$, the current mass of the CEE donor, by 0.028~M$_{\odot}$ to estimate $M_{\rm WD}$, i.e., the mass that the CEE donor would have as a WD if the CEE hadn’t happened. By interpolating their figure, we obtain the following equations:

\begin{eqnarray}
\begin{aligned}
M_{\rm MS} &= 7.42\, M_{\rm WD} - 2.93, &\quad \text{if } M_{\rm WD} < 0.53 \\ 
M_{\rm MS} &= 13.10\, M_{\rm WD} - 5.95, &\quad \text{if } 0.53 \leq M_{\rm WD} < 0.59 \nonumber \\ 
M_{\rm MS} &= 15.41\, M_{\rm WD} - 7.31, &\quad \text{if } 0.59 \leq M_{\rm WD} < 0.61 \nonumber \\ 
M_{\rm MS} &= 36.90\, M_{\rm WD} - 20.36, &\quad \text{if } 0.61 \leq M_{\rm WD} < 0.63 \nonumber \\ 
M_{\rm MS} &= 5.08\, M_{\rm WD} - 0.21, &\quad \text{if } 0.63 \leq M_{\rm WD} < 0.83 \nonumber \\ 
M_{\rm MS} &= 32.26\, M_{\rm WD} - 22.90, &\quad \text{if } 0.83 \leq M_{\rm WD} \leq 0.86 \nonumber
\end{aligned}
\end{eqnarray}

The giant mass is then defined as $M_{1,t} = 0.9~M_{\rm MS} $ for RGBs, and $M_{1,t} = 0.75~M_{\rm MS} $ for AGBs. Finally, to estimate the giant radius, they use the following equation (for solar metallicity):

\begin{equation}
    R_1 = 440~\text{R}_{\odot}\left(\frac{M_{\rm MS}}{\text{M}_{\odot}}\right)^{-0.47}  \left(\frac{M_{\rm1, c}}{0.6~\text{M}_{\odot}}\right)^{5.1}. \nonumber
\end{equation}

Following this procedure, we estimate the pre-CEE parameters for the observations with $M_{\rm MS} < 5~\text{M}_\odot$, which is also the maximum progenitor mass from the AGB simulations listed in Tab.~\ref{tab:A3}. Moreover, in line with the approach of \citet{2019MNRAS.490.2550I}, we do not include error bars when comparing with observations. This choice is justified given that our goal is to perform an order-of-magnitude comparison rather than a statistical analysis.
Figure~\ref{fig:observations}.a presents the plot of $q$ versus $a_{\rm f}/ R_1$ for both observations and simulations. Observations do not follow a linear relationship and consistently lie below the simulation results. Moreover, systems with more massive primary cores (AGBs, red/green circles) disagree with simulations whereas those with lower core masses (RGBs, violet circles) are roughly consistent with predictions. There are several possible explanations for this discrepancy:

\begin{itemize}
\item Observational biases have a preference for detecting short-period systems.
\item Uncertainties in the reconstruction of stellar radii can influence the comparison of observed and simulated data.
\item Most existing simulations are for RGB stars, whereas many observed systems were in the post-AGB phase when CEE happened.
\item Simulations often exclude key physics, such as radiation. 
\item Simulations generally model only the rapid plunge-in phase of the CEE and miss later stages of the event.
\item Orbital decay can continue after the end of the CEE, driven by mechanisms such as magnetic braking or future interactions.
\end{itemize}

We discuss each of these possibilities in the following sections.

\subsection{Observational biases}

Short-period binaries are easier to detect because their full orbits can be observed within a single night. Additionally, many known systems are identified as eclipsing binaries, which are more likely to be found when the orbital period is short, because of the higher eclipse probability. Although some intermediate-period post-CEE systems have recently been identified, they are not as numerous as the short-period population.

As examples of intermediate-period systems with post-AGB giants, recent \textit{Gaia} observations have identified five post-CEE binaries hosting massive WDs ($>1$~M$_{\odot}$) with orbital periods between 18 and 49 days \citep{2024MNRAS.52711719Y}. We include these systems in Figure~\ref{fig:observations}, which shows that their final separations are consistent with the lower-end of simulations.  For this comparison, we adopted the pre-CEE parameters derived by \cite{2024A&A...686A..61B}, since their  $M_{\text{MS}}$ values exceed the 5~$\text{M}_{\odot}$-limit imposed by the reconstruction method described above. An additional illustrative case is  KOI-3278 \citep{2014Sci...344..275K}. Using the pre-CEE parameters from \cite{2024A&A...687A..12B}, its $a_f/R_1$ value is also closer to the simulation predictions than to those of other post-AGB systems in Fig.~\ref{fig:observations}.a.

Those systems do not eliminate the discrepancies, given that current simulations are not able to reproduce post-AGB systems with smaller separations, but increases the number of post-AGB systems that align with the predicted results.

\subsection{Reconstruction errors}

AGB simulations follow the same linear fit as RGB simulations because they compensate for their larger radii by also having larger $a_{\rm f}$. In contrast, post-AGB observations have similar $a_{\rm f}$ to RGB simulations~(see Fig.~\ref{fig:observations}.b) and therefore end up falling well below the expected fit.  Although overestimation of post-AGB radii could partially reduce this tension, reconciling observations with simulations would require shrinking the inferred radii by more than an order of magnitude. Thus, reconstruction errors are unlikely to fully account for the discrepancy.

\subsection{Limited parameter space of AGB simulations}

As discussed above, observations of post-AGB systems tend to have smaller $a_{\rm f}$ than predicted by AGB simulations. Most existing simulations focus on RGB donors, with only a limited number exploring AGB stars. Although available AGB simulations generally predict a larger $a_{\rm f}$, the small sample leaves open the possibility that future simulations may reveal cases with smaller $a_{\rm f}$. If such systems exist, they could bring the simulations closer to agreement with the observed post-AGB population.

\subsection{Absence of physics in simulations} 

It is possible that some physics is missing in simulations. However, this seems to have a larger effect on the unbound fraction of the envelope than on the final separation. For example, \cite{2020A&A...644A..60S} simulated AGB systems with and without recombination energy and found that $a_{\rm f}$ differ by around 30\%, which is less than the impacts of numerics discussed in Sec.~\ref{sec:other_simulations}. Moreover, \cite{2025arXiv250320506L} included radiation transport in their simulations with the same giant as \cite{2022MNRAS.512.5462L} and found similar $a_{\rm f}$. In both cases, the addition of physics has a high impact on the ejection of the envelope but not on $a_{\rm f}$.

\subsection{Orbit decay after the end of the plunge-in, but before the end of CEE}

Another possibility is that most hydrodynamical simulations capture mostly the initial plunge-in phase of CEE. However, the event unfolds in multiple stages, beginning with the loss of dynamical stability and followed by a rapid plunge-in. If the envelope is not fully ejected during this stage, the subsequent evolution is often not entirely modeled. As a result, the final orbital separations in simulations may reflect conditions at the end of the plunge-in rather than the actual post-CEE configuration. If some envelope material remains bound, further orbital decay is expected, and may even result in a merger. This framework helps explain why simulations tend to define an upper boundary for the observed population without a clear gap.

This may also explains why there is a  better agreement between the simulations and post-RGB observations compared to post-AGB systems. Although AGB stars have more weakly bound envelopes, their lower densities envelope may reduce drag forces during the plunge-in, limiting orbital energy loss and leading to larger $a_{\rm f}$ right after the plunge. Those giants may need to keep the orbital decay during the slower spiral-in phase. In contrast, RGB stars, with higher-density envelopes, may extract more orbital energy during the plunge.

\subsection{Orbit decay after the end of CEE}

Although systems flagged as undergoing ongoing mass transfer have been excluded, this does not rule out the possibility that interactions occurred after the end of CEE. Moreover, some degree of orbital decay is expected, driven by mechanisms such as magnetic braking or gravitational wave emission, which can further reduce the separation of post-CEE binaries. Like the previous explanation, this scenario also accounts for why simulations provide an upper boundary for observations.

\section{Moving forward}\label{sec:moving}

The traditional CEE framework typically divides the process into three main phases:
\begin{itemize}
\item (1) unstable accretion;
\item (2) a rapid plunge-in;
\item (3) a slower spiral-in.
\end{itemize}

Hydrodynamical simulations often begin at the onset of phase (2) and extend up to the start of phase (3). Although the exact moment when phase (2) ends and phase (3) begins is somewhat arbitrary, the distinction between the two phases is physically meaningful. During phase (2), the orbit of the secondary around the primary core is generally non-Keplerian and the envelope is highly asymmetric. In phase (3), the system consists of a close binary surrounded by not ejected envelope material.

More importantly, the time it takes for each phase to end is very different. The plunge-in phase typically lasts only months (in our simulations, from a few months to a couple of years), whereas the slower spiral-in phase can persist for decades or even centuries. Consequently, the physical processes relevant to phase (3) may not play a significant role during phase (2). For example, while accretion may occur throughout the event, its impact on orbital decay is less important during the dynamical plunge-in, when orbital energy dominates, but can become dominant during the slower spiral-in. In fact, \cite{2020cee..book.....I} estimated that accretion becomes an important source of energy when the slow spiral-in lasts longer than $\sim$30 years for a $\sim$100~R$_{\odot}$ envelope (and for shorter durations if the envelope radius is larger). Thus, dividing the CEE into distinct phases allows analytical models to include only the physical processes that are relevant in each stage.

We therefore advocate that CEE should treat it as a sequence of distinct phases rather than as a single monolithic event, particularly in 1D or analytical frameworks. In this picture, the resulting orbital evolution for a standard CEE can be expressed as

\begin{equation}\label{eq:phases_a}
a_{\mathrm{obs}} = a_{\mathrm{initial}} - \Delta a_{\mathrm{unst}} - \Delta a_{\mathrm{plunge}} - \Delta a_{\mathrm{spiral}},
\end{equation}

\noindent where each $\Delta a_i$ represents the orbital separation lost during a physically distinct phase: the initial unstable accretion ($\Delta a_{\mathrm{unst}}$), the dynamical plunge-in ($\Delta a_{\mathrm{plunge}}$), and the subsequent slow spiral-in ($\Delta a_{\mathrm{spiral}}$).

A couple 1D descriptions for the plunge phase already exist and can be considered to estimate $\Delta a_{\mathrm{plunge}}$~\citep{2022PhRvD.106d3014T,2024A&A...683A..65B}. In this study, we found an empirical linear correlation that can also be used to estimate the orbital separation following the end of the plunge, leading to:

\begin{equation}
a_{\mathrm{obs}} \simeq (0.31q - 0.011)R_1 - \Delta a_{\mathrm{spiral}}.
\end{equation}

The detailed evolution during the slow spiral-in remains uncertain. In our simulations, we observe continued orbital decay for the 2~$\text{M}_{\odot}$ RGB case, where the separation continues to decrease after the plunge. Unfortunately, our current simulations do not include additional energy sources such as accretion, recombination, or radiation, which, as discussed above, may become critical during this stage. Furthermore, as noted by \citet{2025arXiv250619547B} and \citet{2025A&A...697A..68G}, limitations in numerical resolution and gravitational softening make it difficult to model the long-term torque evolution during this phase, and thus the complete orbital decay, up to the end of the spiral-in phase.

A few analytical \added{and 1D} models for the slower spiral-in phase have also been proposed. For instance, \cite{2017MNRAS.470.1788C} suggested that episodic plunges, in which the system undergoes successive partial plunges, each generating drag onto the close binary. Thus, decreasing the orbital separation and releasing enough energy to unbind part of the envelope before drag subsides again. Another possibility involves a circumbinary disk formed by the bound material accreting onto the central binary. Accretion from this disk modifies the binary’s orbital separation while driving bipolar outflows that expel the envelope~\citep{2023ApJ...955..125T}. Finally, \cite{2022ApJ...937L..42H} proposed that late-time stable mass transfer may occur in red supergiants when the outer convective envelope is ejected during the plunge, while the more tightly bound radiative layer remains, allowing the remnant to refill its Roche lobe. Thus, there is still not a consensus on how the slow spiral happens and whether the main mechanism is the same for all giant types.

Observations of post-CEE systems support the idea that not all systems go through every phase. Wide post-CEE systems, such as V1379~Aql, may experience the unstable accretion phase but not a classical plunge \citep{2019MNRAS.490.2550I}, while systems such as KOI-3278 may have stopped evolving right after the plunge without undergoing a long spiral-in phase. Treating CEE as a multi-phase process naturally allows different systems to follow distinct evolutionary tracks. In such cases, Eq.~\ref{eq:phases_a} can be modified to exclude phases as appropriate.

Although a multi-phase analytical model introduces additional complexity, it is physically motivated and can improve the predictive power of CEE models. The single-$\alpha$ energy formalism has long been recognized as an incomplete physical description, as it neglects various energy sources and sinks such as recombination, radiation, and accretion. In fact, some alternative versions of the formalism have already been proposed to account for extra energy sources, for example, \cite{2022MNRAS.512.5462L}. 

However, despite its shortcomings, the single-$\alpha$ formalism remains practically advantageous because it does not require detailed knowledge of CEE’s evolution. Since CEE events are rarely directly observed in real time (except for mergers ending in LRNe), this simplicity is nontrivial and has made the single-$\alpha$ approach particularly useful for population synthesis studies. In contrast, a multi-phase description requires constraining the behavior of each phase, which remains challenging. In the long term, however, modeling CEE as a sequence of distinct phases offers a promising and physically grounded path forward, even though developing a complete framework will require substantial observational, theoretical, and numerical input. In this context, our results provide an empirical fit for the final separation after the dynamical plunge-in phase, an important step toward building a multi-phase framework for CEE.

\section{Summary}
\label{sec:summary}

Below, we summarize the main findings and discussion of this work:

{\bf(i) Our simulations show a roughly linear correlation between final separation and mass ratio.}

Although the true relation may not be strictly linear, this approximation holds very well throughout the explored range and provides a practical tool for predicting CEE outcomes during the plunge-in phase. This trend is not easily explained by the energy formalism, given that the envelope is still partially bound by the end of simulations.

{\bf (ii) Other simulations with RGB, AGB and supergiant stars agree with the linear trend.} 

Although the true slope may vary between different giant masses and evolutionary stages and other physical parameters, these differences are typically smaller than the differences introduced by numerical methods. As a result, a single linear correlation provides a reliable estimate of the final separation for $q\geq 0.15$.

 {\bf(iii) Observations mostly do not match the linear fit.}
 
With some exceptions, AGB observations of post-CEE binaries are found well below the linear fit predicted by simulations. Most simulations track mostly the plunge-in phase, during which rapid orbital decay occurs, and only the beginning of the slower spiral-in phase. However, in reality, orbital decay likely continues during the slower spiral-in phase, which may explain this mismatch. This highlights the importance of modeling CEE as a series of phases for an accurate comparison with observations.

 {\bf(iv) An analytical model for CEE may require dividing the event into phases.}

While CEE is traditionally divided into at least three phases, the importance, duration, and sequence of these phases can vary from system to system. Some binaries may eject most of the envelope during the plunge, while others continue to a gradual orbital decay and envelope ejection in a slow spiral-in. Modeling CEE in a analytical framework that is modular would allow each system’s unique evolutionary path to be captured. In this sense, the linear correlation found in Sec.~\ref{sec:other_simulations} can be used as an approximation for the separation right after the plunge. 

\begin{acknowledgments}
SVB thanks the anonymous referee for the comments, Philip Chang for the numerous discussions that significantly contributed to this paper, and Logan Prust for the insightful comments. SVB also acknowledges support from the NASA ATP program (NNH23ZDA001N-ATP) and the College of Letters and Science at UWM through the Chancellor's Graduate Student Award and the Research Excellence Award.

\end{acknowledgments}

\software{MANGA, MESA,
YT \citep{2011ApJS..192....9T}, Astropy~\citep{2013A&A...558A..33A, 2018AJ....156..123A, 2022ApJ...935..167A}.
          }

\section*{Data Availability}

The following data are available at \url{https://doi.org/10.5281/zenodo.17852204}:

\begin{itemize}
\item 13 \texttt{.dat} files containing the time evolution of the final separation (one for each simulation),
\item 13 \texttt{.dat} files containing the time evolution of the unbound mass fraction (one for each simulation),
\item Two \texttt{.dat} files with the simulations presented in Tables~\ref{tab:A3} and~\ref{tab:new_sims},
\item A Python script (\texttt{generate\_fig.py}) to generate Figure~\ref{fig:observations}.
\end{itemize}

The data to reproduce the numerical simulations and Fig~\ref{fig:corotation} is available on reasonable request. All other Figures can be reproduced with the information available online. To evolve the star with MESA, we used version r23.05.1 and the example "1M\_pre\_ms\_to\_wd" from the test suite (adjusting the ZAMS mass for the 2M$_{\odot}$ RGB case).

\begin{table*}
\caption{Summary of CEE simulations with AGB/RGBs used in Fig.~\ref{fig:all_correlation}. Simulations were taken from Tab. A3 of \citet{2019MNRAS.490.2550I}. The masses are in M$_\odot$ and distances in R$_\odot$.}
\label{tab:A3}
\vspace{-2 em}
\begin{center}
\begin{tabular}{cccccccc|cccccccc}
\hline
Type & $q$ & $M_{1,\mathrm{t}}$ & $M_{1,\mathrm{c}}$ & $R_1$ & $a_{\rm i}/R_1$ & $a_{\rm f}$ & Ref. &
Type & $q$ & $M_{1,\mathrm{t}}$ & $M_{1,\mathrm{c}}$ & $R_1$ & $a_{\rm i}/R_1$ & $a_{\rm f}$ & Ref. \\
\hline

 AGB & 0.08 & 5 & 1.0 & 200 & 1.4 & 4.4 & (1) &RGB & 0.50 & 1.98 & 0.38 & 52 & 1.0 & 3.6 &(9)\\
AGB & 0.12 & 5 & 1.0 & 200 & 1.4 & 4.8 & (1)& RGB & 0.50 & 1.96 & 0.37 & 48 & 1.0 & 8.0 & (10)\\
AGB & 0.12 & 5 & 0.94 & 354 & 1.5 & 8.9 &(1) &RGB & 0.57 & 1.05 & 0.36 & 31 & 2.0 & 9.0 &(11) \\
AGB & 0.13 & 3 & 0.7 & 200 & 1.4 & 4.4 &(1) & RGB & 0.27 & 1.18 & 0.36 & 60 & 2.0 & 3.2 &(12) \\
AGB & 0.13 & 3 & 0.7 & 200 & 1.4 & 4.7 &(1) & RGB & 0.31 & 1.18 & 0.36 & 60 & 2.1 & 3.7 &(12) \\ 
RGB & 0.18 & 4 & 0.7 & 66 & 1.6 & 1.0 & (2)&RGB & 0.34 & 1.18 & 0.36 & 60 & 2.1 & 3.5 & (12)\\
RGB & 0.35 & 1 & 0.45 & 243 & 1.3 & 21 & (3)& RGB & 0.23 & 1.38 & 0.36 & 57 & 2.0 & 2.5 &(12) \\
RGB & 0.10 & 1 & 0.45 & 221 & 1.3 & 33 & (3)&RGB & 0.26 & 1.38 & 0.36 & 57 & 2.0 & 2.8 &(12)\\
RGB & 0.18 & 2 & 0.45 & 177 & 1.3 & 19 &(3) & RGB & 0.29 & 1.38 & 0.36 & 57 & 2.0 & 3.0 &(12)\\
RGB & 0.11 & 0.88 & 0.39 & 83 & 1.0 & 3.1 &(4)&RGB & 0.20 & 1.59 & 0.36 & 50 & 1.9 & 1.7 &(12)\\
RGB & 0.17 & 0.88 & 0.39 & 83 & 1.0 & 3.4 &(4)&  RGB & 0.23 & 1.59 & 0.36 & 50 & 2.0 & 1.8 &(12)\\
RGB & 0.34 & 0.88 & 0.39 & 83 & 1.0 & 7.3 &(4)&RGB & 0.25 & 1.59 & 0.36 & 50 & 2.0 & 2.1 &(12)\\
RGB & 0.68 & 0.88 & 0.39 & 83 & 1.0 & 12 &(4)& RGB & 0.18 & 1.8 & 0.36 & 41 & 1.9 & 1.2 &(12)\\
RGB & 1.02 & 0.88 & 0.39 & 83 & 1.0 & 17 &(4)&RGB & 0.20 & 1.8 & 0.36 & 41 & 1.9 & 1.3 &(12)\\
RGB & 0.05 & 1.97 & 0.39 & 66 & 1.0 & 2.2 &(4)&RGB & 0.22 & 1.8 & 0.36 & 41 & 2.0 & 1.4 &(12)\\
RGB & 0.08 & 1.97 & 0.39 & 66 & 1.0 & 1.8 &(4)& RGB & 0.10 & 2 & 0.33 & 44 & 1.3 & 2.1 &(3)\\
RGB & 0.15 & 1.97 & 0.39 & 66 & 1.0 & 1.6 &(4)& RGB & 0.24 & 1.5 & 0.32 & 26 & 2.0 & 0.91&(13) \\
RGB & 0.30 & 1.97 & 0.39 & 66 & 1.0 & 1.8 &(4)&  RGB & 0.27 & 1.2 & 0.32 & 29 & 2.0 & 1.4 &(12)\\
RGB & 0.46 & 1.97 & 0.39 & 66 & 1.0 & 2.8 &(4)& RGB & 0.30 & 1.2 & 0.32 & 29 & 2.1 & 1.5&(12) \\
RGB & 0.30 & 1.97 & 0.39 & 66 & 1.0 & 0.67 &(4)&RGB & 0.33 & 1.2 & 0.32 & 29 & 2.1 & 1.4 &(12)\\
RGB & 0.30 & 1.97 & 0.39 & 66 & 1.0 & 0.87 &(4)& RGB & 0.23 & 1.4 & 0.32 & 28 & 2.0 & 1.1 &(12)\\
RGB & 0.11 & 0.88 & 0.39 & 83 & 1.0 & 6.1 &(5)& RGB & 0.26 & 1.4 & 0.32 & 28 & 2.0 & 1.1&(12) \\
RGB & 0.17 & 0.88 & 0.39 & 83 & 1.0 & 7.3 &(5)& RGB & 0.29 & 1.4 & 0.32 & 28 & 2.0 & 1.2 &(12)\\ 
RGB & 0.34 & 0.88 & 0.39 & 83 & 1.0 & 11 &(5)&  RGB & 0.20 & 1.6 & 0.32 & 26 & 1.9 & 0.87&(12) \\
RGB & 0.68 & 0.88 & 0.39 & 83 & 1.0 & 21 &(5)&   RGB & 0.23 & 1.6 & 0.32 & 26 & 2.0 & 0.91&(12) \\
RGB & 1.02 & 0.88 & 0.39 & 83 & 1.0 & 27 &(5)&  RGB & 0.23 & 1.6 & 0.32 & 31 & 1.6 & 0.93&(12) \\
RGB & 0.68 & 0.88 & 0.39 & 100 & 3.0 & 20 &(6)&RGB & 0.25 & 1.6 & 0.32 & 26 & 2.0 & 0.96&(12) \\
RGB & 0.68 & 0.88 & 0.39 & 87 & 2.5 & 30 &(7)& RGB & 0.18 & 1.8 & 0.32 & 16 & 1.9 & 0.43 &(12)\\
RGB & 0.68 & 0.88 & 0.39 & 91 & 2.4 & 28 &(7)&  RGB & 0.20 & 1.8 & 0.32 & 16 & 1.9 & 0.48&(12) \\ 
RGB & 0.68 & 0.88 & 0.39 & 93 & 2.3 & 28 &(7)& RGB & 0.22 & 1.8 & 0.32 & 16 & 2.0 & 0.53&(12) \\
RGB & 0.50 & 1.98 & 0.38 & 49 & 1.0 & 4.9 &(8)& RGB & 0.35 & 1 & 0.28 & 22 & 1.3 & 1.8 &(3)\\
RGB & 0.50 & 1.98 & 0.38 & 52 & 1.0 & 3.2 &(9)& RGB & 0.45 & 1 & 0.28 & 18 & 1.3 & 2.4 &(3)\\

\end{tabular}
\end{center}
\vspace{-0.5 em}
\small{References: (1) \citet{1998ApJ...500..909S}; (2) \citet{1996ApJ...471..366R}; (3) \citet{2000ApJ...533..984S}; (4) \citet{2018MNRAS.477.2349I}; (5) \citet{2012ApJ...744...52P}; (6)  \citet{2017MNRAS.464.4028I}; (7) \citet{2019MNRAS.484..631R}; (8) \citet{2016ApJ...816L...9O};
(9) \citet{2019MNRAS.486.5809P}; (10) \citet{2018MNRAS.480.1898C}; (11) \citet{2012ApJ...746...74R}; (12) \citet{2016MNRAS.460.3992N};  (13) \citet{2015MNRAS.450L..39N}.}
\end{table*}

\begin{table*}
\caption{Summary of CEE simulations with AGB, RGB and Supergiants~(SGs) used in Fig.~\ref{fig:all_correlation} that are not present in \cite{2019MNRAS.490.2550I}. The masses are in M$_\odot$ and distances in R$_\odot$. }
\label{tab:new_sims}
\vspace{-2 em}
\begin{center}
\begin{tabular}{cccccccc|cccccccc}
\hline
Type & $q$ & $M_{1,\mathrm{t}}$ & $M_{1,\mathrm{c}}$ & $R_1$ & $a_{\rm i}/R_1$ & $a_{\rm f}$ & Ref. &
Type & $q$ & $M_{1,\mathrm{t}}$ & $M_{1,\mathrm{c}}$ & $R_1$ & $a_{\rm i}/R_1$ & $a_{\rm f}$ & Ref. \\
\hline
AGB & 0.25 & 0.97 & 0.545 & 173 & 1.20 & 17 &(1) & RGB &0.68 &0.88 &0.39 &88 &1.14 &14.5& (10) \\
AGB & 0.50 & 0.97 & 0.545 & 173 & 1.36 & 34 &(1) & RGB &0.2 &1.80 &0.32 &17 &1.85 &0.49& (10)\\
AGB & 0.75 & 0.97 & 0.545 & 173 & 1.49 & 55 &(1) & RGB &0.2 &1.80 &0.32 &16 &1.96 &0.51 &(10)\\
AGB & 0.25 & 0.97 & 0.545 & 173 & 1.20 & 22 &(1) & RGB &0.2 &1.80 &0.313 &16 &1.96 &0.44 &(10)\\
AGB & 0.50 & 0.97 & 0.545 & 173 & 1.36 & 41 &(1) & RGB &0.2 &1.80 &0.313 &15 &2.09 &0.37& (10) \\
AGB & 0.75 & 0.97 & 0.545 & 173 & 1.49 & 69 &(1) & RGB & 0.05 & 1 & 0.38 & 78 & 1 & 3.5  &(11) \\
AGB & 0.25 & 0.97 & 0.545 & 173 & 1.20 & 21 &(2) &  RGB & 0.3 & 1 & 0.38 & 78 & 1 & 7.8  &(11) \\
AGB & 0.50 & 0.97 & 0.545 & 173 & 1.36 & 37.8&(2) &  RGB & 0.6 & 1 & 0.38 & 78 & 1 & 14.4 &(11) \\
AGB & 0.75 & 0.97 & 0.545 & 173 & 1.48 & 54.2 &(2) & RGB & 0.9 & 1 & 0.38 & 78 & 1 & 22.7  &(11) \\
AGB &0.35 &1.7 &0.56 &250 &2.2 &32.7& (3) & RGB & 0.5 & 2 & 0.37 & 53 & 1 & 3.8  &(11) \\
AGB &0.35 &1.7 &0.56 &250 &2.2 &25.9& (3) &RGB & 0.25 & 2 & 0.37 & 53 & 1 & 3.5  &(11) \\
AGB &0.35 &1.7 &0.56 &250 &2.2 &20.6 &(3) &  RGB & 0.5 & 2 & 0.37 & 53 & 1 & 6.9  &(11) \\
AGB &0.35 &1.7 &0.56 &250 &2.2 &25.8& (3)  & RGB & 0.75 & 2 & 0.37 & 53 & 1 & 11.2  &(11) \\
AGB & 0.01 & 0.77 & 0.47 & 173 & 0.77 & 12.4  & (4) &  RGB & 1 & 2 & 0.37 & 53 & 1 & 16.7 &(11) \\  
AGB & 0.10 & 0.77 & 0.47 & 173 & 0.95 & 11.6  &(4) &  RGB & 0.25 & 2 & 0.37 & 53 & 1 & 2.8  &(11) \\
AGB & 0.10 & 0.77 & 0.47 & 173 & 0.95 & 10.4  &(4) &   RGB & 0.5 & 2 & 0.37 & 53 & 1 & 5.7  &(11) \\
AGB & 0.06 & 0.77 & 0.47 & 173 & 0.89 & 7.5  &(4) &RGB & 0.75 & 2 & 0.37 & 53 & 1 & 9.1  &(11)  \\
AGB & 0.04 & 0.77 & 0.47 & 173 & 0.85 & 14.2  &(4) & RGB & 1 & 2 & 0.37 & 53 & 1 & 12.4  &(11)  \\
AGB &0.55 &1.78 &0.53 &122.2 &1.01 &15.5 &(5) & SG & 0.15 &  9.61 & 2.96 & 438 & 1.1 & 15.12   & (12) \\  
AGB & 0.35 & 1.7 &0.56 & 260 & 2.12 &21 & (6) & SG &0.52 &  9.61 & 2.96 & 438 & 1.14 & 47.48   & (12) \\
AGB & 0.35 & 1.7 &0.56 & 260 & 2.12 &23 & (6)&  SG &0.25 & 12.0 & 3.84 & 619 & 1.60 & 34.3 & (13) \\
AGB & 0.35 & 1.7 &0.56 & 260 & 2.12 &26 & (6) & SG &0.25 & 12.0 & 3.84 & 619 & 1.60 & 37.6 & (13) \\
AGB & 0.35 & 1.7 &0.56 & 260 & 2.12 &32 & (6) & SG &0.25 & 12.0 & 3.84 & 619 & 1.60 & 38.5 & (13) \\
AGB & 0.35 & 1.7 &0.56 & 260 & 2.12 &33 & (6) & SG &0.25 & 12.0 & 3.84 & 619 & 1.60 & 36 & (13) \\
AGB & 0.35 & 1.7 &0.56 & 260 & 2.12 &40 & (6) & SG &0.25 & 12.0 & 3.84 & 619 & 1.60 & 40.9 & (13) \\
AGB & 0.16 & 3.7 &0.72 & 330 & 1.93 &8.2 & (6) & SG &0.25 & 12.0 & 3.84 & 619 & 1.60 & 41 & (13) \\
AGB & 0.16 & 3.7 &0.72 & 330 & 1.93 &7.9 & (6)& SG &0.25 & 12.0 & 3.84 & 619 & 1.60 & 44.3 & (13) \\
AGB & 0.16 & 3.7 &0.72 & 330 & 1.93 &9.4 & (6) & SG &0.25 & 12.0 & 3.84 & 619 & 1.60 & 33.1 & (13) \\
AGB & 0.16 & 3.7 &0.72 & 330 & 1.93 &9.9 & (6)& SG &0.25 & 12.0 & 3.84 & 619 & 1.60 & 30.2 & (13) \\
AGB & 0.16 & 3.7 &0.72 & 330 & 1.93 &12 & (6) & SG &0.53 &  9.4 & 2.97 & 395 & 1.27 & 37   & (14) \\
AGB & 0.16 & 3.7 &0.72 & 330 & 1.93 &10 & (6) & SG &0.15 &  9.4 & 2.97 & 395 & 1.21 & 14   & (14) \\
AGB & 0.35 & 1.7 &0.56 & 260 & 2.12 &42 & (7) & SG &0.5 &  10 & 2.97 & 438 & 1.14 & 46   & (15) \\
AGB & 0.16 & 3.7 &0.72 & 343 & 1.86 &11 & (7) & SG & 0.25 & 12.0 & 3.84 & 618 & 1.60 & 37.2 & (16) \\
RGB & 0.5 & 2 & 0.37 & 52 & 1 & 5  &(8)  & SG &0.25 & 12.0 & 3.84 & 618 & 1.60 & 44.8 & (16) \\
RGB &0.5 &1.96 &0.37 &48.1 &1.02 &7 &(5)  & SG &0.25 & 12.0 & 3.84 & 618 & 1.60 & 44.7 & (16) \\
RGB &0.5 &1.96 &0.366 &48.1 &1.02& 3.4& (9) & SG &0.25 & 12.0 & 3.84 & 618 & 1.60 & 38.6 & (16) \\
RGB &0.5 &1.96 &0.366 &48.1 &1.02 &3.5 &(9)& SG &0.25 & 12.0 & 3.84 & 618 & 1.60 & 42.1 & (17) \\
RGB &0.68 &0.88 &0.39 &81 &1.23 &12.5& (10)  & SG &0.25 & 12.0 & 3.84 & 618 & 1.60 & 39.4 & (17) \\

\hline
\end{tabular}
\end{center}
\vspace{-0.5 em}
\small{References: (1) \citet{2020A&A...644A..60S}; (2) \citet{2022A&A...660L...8O}; (3) \citet{2022MNRAS.517.3181G}; (4) \citet{2020A&A...642A..97K}; (5) \citet{2020MNRAS.495.4028C}; (6) \citet{2024MNRAS.527.9145G}; (7) \citet{2024MNRAS.533..464B}; (8) \citet{2023MNRAS.526.5365V}; (9) \citet{2024MNRAS.528..234C}; (10) \citet{2020MNRAS.494.5333R}; (11) This work; (12) \citet{2022A&A...667A..72M}; (13) \citet{2022MNRAS.512.5462L}; (14) \citet{2024A&A...691A.244V}; (15) \citet{2025A&A...698A.133V}; (16) \citet{2022MNRAS.516.4669L} ; (17) \citet{2025A&A...699A.274L}.}
\end{table*}

\bibliography{sample7}{}
\bibliographystyle{aasjournalv7}



\end{document}